\documentclass{article}

\usepackage{arxiv}
\usepackage[utf8]{inputenc} % allow utf-8 input
\usepackage[T1]{fontenc}    % use 8-bit T1 fonts
\usepackage{hyperref}       % hyperlinks
\usepackage{url}            % simple URL typesetting
\usepackage{booktabs}       % professional-quality tables
\usepackage{amsfonts}       % blackboard math symbols
\usepackage{nicefrac}       % compact symbols for 1/2, etc.
\usepackage{microtype}      % microtypography
\usepackage{lipsum}
\usepackage{wrapfig}
\usepackage{lscape}
\usepackage{rotating}
\usepackage{epstopdf}
\usepackage{amsmath,scalerel}
\usepackage[export]{adjustbox}
\usepackage{subfigure}
\usepackage{caption}
\usepackage{bm}
\usepackage{amsfonts}
\usepackage{amssymb}
\usepackage{xcolor,colortbl}
\usepackage{multirow}
\usepackage{multicol}

\usepackage{amsthm}
\usepackage{mathtools}
\usepackage{tikz}
\usepackage{enumerate}
\usepackage{comment}
\usepackage{mathtools}

\usepackage{color}
\usepackage{graphicx}
\usepackage{verbatim}
\usepackage{amsmath}
\usepackage{stmaryrd}
\usepackage{booktabs}
\usepackage{lineno}
\newcommand\myeq{\stackrel{\mathclap{\normalfont\mbox{?}}}{=}}
\title{Trustworthy and Privacy-Aware Sensing \\ for Internet of Things}

\author{
  Ihtesham Haider \\
  Institute of Networked and Embedded Systems\\
  Alpen-Adria-Universit{\"a}t, Klagenfurt Austria\\
  \texttt{ihtesham.hdr@gmail.com} \\
   \And
 Bernhard Rinner \\
  Institute of Networked and Embedded Systems\\
  Alpen-Adria-Universit{\"a}t, Klagenfurt Austria\\
  \texttt{bernhard.rinner@aau.at} \\
}

\begin{document}
\maketitle
\sloppy
\begin{abstract}

The Internet of Things (IoT) is considered as the key enabling technology for smart services. Security and privacy are particularly open challenges for IoT applications due to the widespread use of commodity devices. This work introduces two hardware-based lightweight security mechanisms to ensure sensed data trustworthiness (i.e., sensed data protection and sensor node protection) and usage privacy of the sensors (i.e., privacy-aware reporting of the sensed data) for centralized and decentralized IoT applications.
Physically unclonable functions (PUF) form the basis of both proposed mechanisms. To demonstrate the feasibility of our PUF-based approach, we have implemented and evaluated PUFs on three platforms (Atmel 8-bit MCU, ARM Cortex M4 32 bit MCU, and Zynq7010 SoC) with varying complexities.  We have also implemented our trusted sensing and privacy-aware reporting scheme (for centralized applications) and secure node scheme (for decentralized applications) on a visual sensor node comprising an OV5642 image sensor and a Zynq7010 SoC.  Our experimental evaluation shows a low overhead wrt.~latency, storage, hardware, and communication incurred by our security mechanisms.
\end{abstract}

%\noindent \textbf{Keywords} Trusted sensing, privacy protection, physically unclonable functions, non interactive witness indistinguishable proofs, Internet of things

\section{Introduction}% Introduction

The future smart world involves a living where people will be automatically and collaboratively served by smart devices and smart spaces interconnected via the Internet of Things (IoT).  IoT applications collect data from various data sources. This data is used for intelligence extraction using machine learning models deployed on cloud and/or edge computing infrastructure. The actionable insight obtained from the intelligence extraction is offered as service to end users but also provides resource efficiency, data knowledge and automated decision-making processes to enterprises. 

Numerous smart services are being conceptualized, researched, prototyped, tested and commercially used today. For instance, smartphones embedded with rich sensing capabilities have enabled navigation~\cite{farrell1999global}, m-commerce~\cite{Fueled}, natural-disaster detection and warning systems~\cite{clayton2012community}, environmental monitoring~\cite{carrapetta2010haze}, and citizen journalism~\cite{das2010prism}. With wearable health sensors it is now possible to monitor the blood sugar level and heart pace~\cite{schoenfeld2004remote}, provide assisted living for elderly patients with chronic diseases~\cite{ecaalyx}, and document daily sports activities of individuals~\cite{denning2009balance}. Smart vehicles have enabled autonomous driving~\cite{AD}, cooperative collision avoidance~\cite{hafner2013cooperative}, remote wireless diagnosis of vehicles~\cite{lightner2003wireless}, and traffic flow optimization~\cite{varaiya1993smart}. Likewise, smart homes~\cite{cook2003mavhome}, and safe cities~\cite{ballesteros2012safe} are enabling smart spaces that are intelligent, resource efficient, and secure.

\begin{figure*}
\centering
\includegraphics[width=6in]{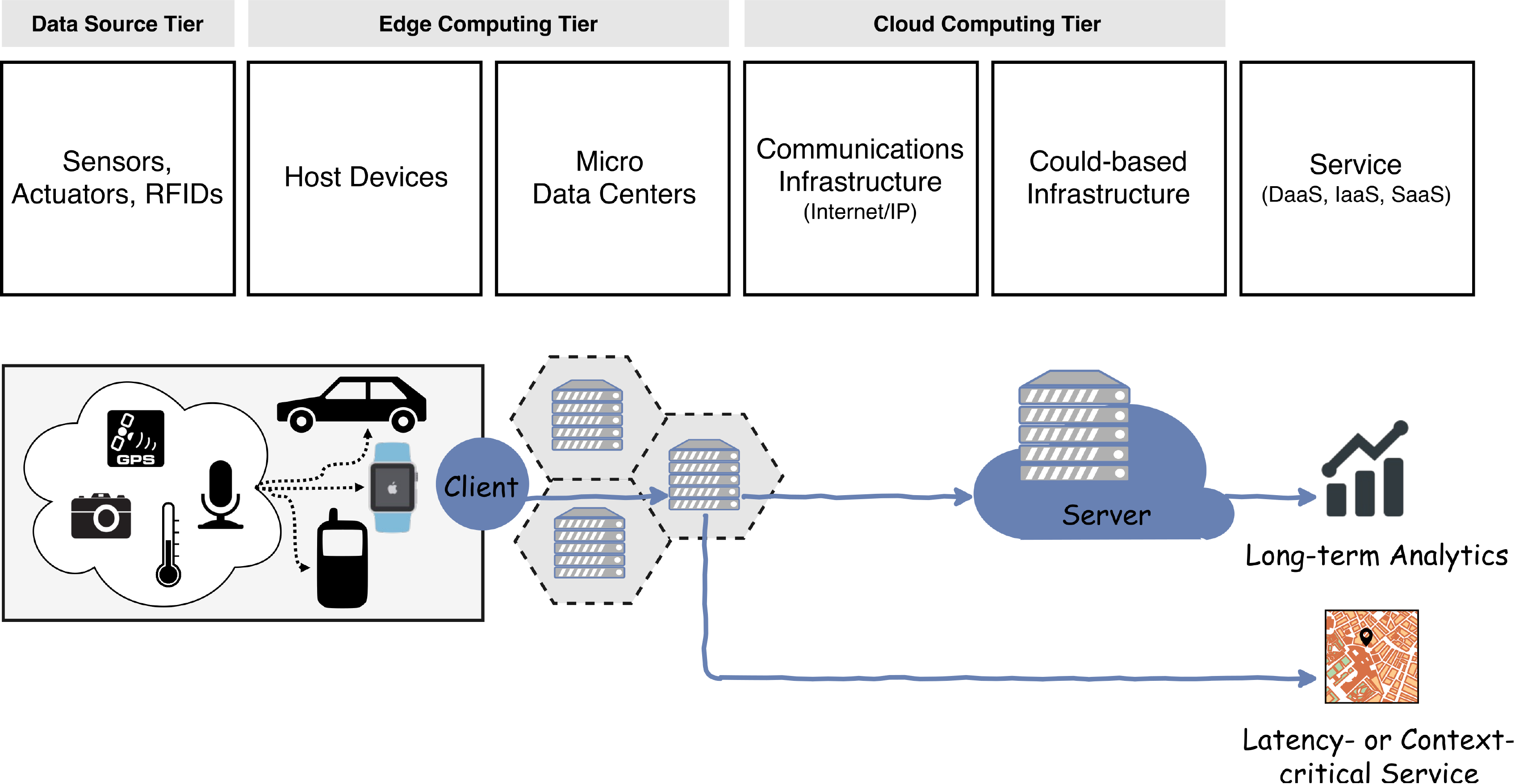}
\caption{A generic infrastructure for IoT applications}
\label{fig:main}
\end{figure*} 

The physical infrastructure of today's IoT applications, as depicted in Fig.~\ref{fig:main}, can be divided into three tiers: data source, edge computation, and cloud computation. The data source tier includes everything that generates data. \textit{Sensors} are the largest and the most common source of data in IoT applications. Other sources include RFIDs, machine logs, social media feeds and event sources. The edge computing tier comprises \textit{host devices} and \textit{micro data centers}, which are responsible for running data  processing pipelines, handling network switching, routing, load balancing and security. A host device can either be a commodity device with computing, storage and communication resources such as a smartphone, a computer, a gateway router, a smart vehicle ECU or a processing platform solely dedicated to the attached data sources. Micro data centers host virtualization infrastructure which runs cloud services closer to the data sources. These data centers are distributively located, for example at cellular base-station sites. The cloud computing tier comprises a centralized pool of computing, storage and communication resources, which offers data management, analytics, software or hardware platform or combination of these as-a-Service (aaS).

The infrastructure of Fig.~\ref{fig:main} encompasses three layers of abstraction: technology, middleware, and application layers~\cite{atzori2010internet}. The technology layer is comprised of sensing, identification, computing and communication resources. The middleware is a software layer or a set of sub-layers, that resides between the technology and the application layer. The middleware hides the technology-level details from the application programmers thereby simplifying the application development process. The application layer is the top most layer exporting all the system's functionality to the end-users by exploiting the functionality of the middleware, standard web interfaces, and protocols. 

\subsection{Security and Privacy Threats}% 

This work addresses two security threats for IoT applications: sensed data pollution and personal privacy leakage.

Sensed data pollution is a major threat posed to IoT applications whereby malicious users or a third-party adversaries contribute manipulated or fabricated sensed data to pollute the application database~\cite{saroiu2010sensor}. An adversary can exploit a number of security vulnerabilities present in the infrastructure of Fig.~\ref{fig:main} to mount this attack. First, the data source layer is mainly comprised of commodity devices embedded with a multitude of resource constrained sensors connected to a host processor. Sensitive data captured by these sensors do not carry any security guarantees. These sensors rely on a resource-rich host device for processing and reporting the sensed data to a micro data center (or a cloud server). Host devices are the commodity devices running a thick, vulnerable software stack~\cite{Gdata2017,saroiu2010sensor}. 
As a result, today it is trivial to manipulate or fabricate sensors' readings in these applications  by exploiting bugs (e.g., \textit{Master Key}~\cite{MasterKey} and \textit{Fake ID}~\cite{FakeID}) in the software stack (e.g., OS) running on the commodity devices. For instance, location readings from a smartphone GPS sensor can be modified to obtain illegitimate access to a location-based service~\cite{liu2012software}, video frames from a surveillance camera can be manipulated to hide or fake an event~\cite{Winkler_COMPSURV2014}, and patient's blood sugar level measured by wearable sensors can be manipulated to stop the insulin pump~\cite{liu2012software}. Second, the use of public infrastructure for communications (Internet) and storage (public cloud storage servers) further increases the threat surface area of these applications. Third, due to the open and ubiquitous nature of the infrastructure, certain elements such as sensors may not be protected against physical attacks. Physically damaged sensors are another potential source of data pollution attacks. Consequently, any service based on this data lacks trust.

The second threat addressed by this work is leakage of personal privacy. The IoT applications collect and process information from almost every aspect of our daily lives, e.g., our private data (photos, medical reports), our routines, habits and preferences (transportation, shopping, political and religious views), our critical infrastructures (energy, emergency systems).  By linking individual data points obtained from wearable, personal devices (e.g., smartphones or cars), and private spaces (e.g., home security, assisted living or baby monitoring applications) one can construct the personal profiles of the individuals revealing their sensitive personal information such as home and workplace, contact details, health-status, religious orientation, political affiliations, current and future locations etc.

\subsection{Contributions}

The goals of this work are twofold:  First, we identify and apply security mechanisms to ensure effective and verifiable trustworthiness of the sensed data, collected from vulnerable commodity devices in open networks such as IoT. We adopt the definition of trustworthy sensed data by Liu et al.~\cite{liu2012software} (i.e., the data carrying integrity, authenticity, and freshness guarantees) and ensure the trustworthiness of sensed data in the IoT applications. Second, certain IoT applications may require these sensors to capture sensitive personal information about the individuals.  We incorporate by design personal privacy protection mechanisms which allows individuals to submit sensed data in a privacy-aware manner.

The diverse nature of the IoT applications impose varying requirements on the data source tier. We categorize these applications into two groups: (i) applications attributed by centralized processing, i.e., in these applications, raw sensed data is collected at the server side (micro data center or cloud sever) for processing and (ii) applications attributed by distributed processing, i.e., sensed data is processed locally on the sensor nodes. 

This manuscript extends our preliminary work on trusted sensing~\cite{haidersecuring, haider2017private} and comprehensively introduces the concepts of trusted sensing and secure camera nodes in a holistic IoT setting. In particular, we first expand trusted sensing~\cite{haidersecuring} by addressing the personal privacy leakage caused by the incorporation of trusted sensors into smart devices such as smart phones. Second, we extend secure camera nodes~\cite{haider2017private} for visual monitoring applications by exploring various PUF sources as root of trust for secure sensor node implementation. Furthermore, both concepts are evaluated using real word application scenarios. Overall, the main contributions of this work can be summarized as follows:

\begin{itemize}

\item First, we present a trusted sensing concept for centralized IoT applications. This concept exploits lightweight security circuits called physically unclonable functions (PUFs) to extract a unique CMOS fingerprint of the sensor. The fingerprint in combination with lightweight security mechanisms ensure non-repudiation (i.e., integrity, authenticity and freshness) on sensor readings.  On-chip PUFs assist to detect  hardware tampering of the sensor.

\item Second, for the centralized IoT applications, we perform anonymization of the sensed data from trusted sensors on the host device using non-interactive witness indistinguishable proofs to ensure privacy-aware submission of the sensed data to the IoT applications.

\item Third, we present a secure node architecture for applications that require processing of sensed data locally on the sensor nodes. The architecture, implemented as system-on-chip, derives the security keys from  sensor's PUF-based CMOS fingerprint. Integrity, authenticity, confidentiality, freshness and access authorization of the sensed data is protected using an encrypt-then-sign technique. Secure boot of the SoC ensures integrity, authenticity and unclonability of the node's firmware. Hardware tampering can be detected due to the tamper evidence property of the on-chip PUF.

\item Fourth, we evaluate both  mechanisms using two case studies. The trusted sensing and anonymization of data from trusted sensors are evaluated using a participatory sensing scenario where a secure node approach is evaluated using a private space monitoring scenario.  A trusted image sensor and a secure camera node are implemented using Zynq7010 SoC and OV5642 5MP image sensor as platform and latency, hardware, storage  and communication overhead incurred by both the approaches is evaluated. To demonstrate the feasibility of PUF-based approach, we also implemented and evaluated PUF on three platforms (Atmel 8-bit MCU, ARM Cortex M4 and Zynq7010 SoC) of varying complexities that are ideally suited as sensing platforms for a  broad range of sensors. 

\end{itemize}

The remainder of the paper is organized as follows: Section~\ref{SOTA} presents the state of the art technologies for ensuring sensed data trustworthiness and sensors' usage privacy in IoT applications. Section~\ref{overview} provides an overview of the employed approach. We present the details of our \textit{trusted sensing and privacy-aware reporting} scheme for centralized IoT applications in Section~\ref{sec:TrustedSensing} and the \textit{secure node} scheme for decentralized applications in Section~\ref{securenode}. We evaluate both schemes in Section~\ref{implementation} and discuss relevant security and privacy properties. Section~\ref{conc} concludes the paper.

\section{Related Work}\label{SOTA}

This section discusses the relevant available work on sensed data trustworthiness and personal privacy protection mechanisms in IoT scenarios. 

The threat surface area of the IoT applications necessitates the protection of data closer to the data source(s).  Securing sensor nodes in the IoT scenario entails data security, node security and usage privacy~\cite{Winkler_COMPSURV2014}.  Data protection is typically implemented in firmware. Any modification in the underlying hardware can completely bypass the data protection. Therefore, node security is an essential requirement for data protection. Moreover, given the ubiquitous and unprotected nature of IoT infrastructure (especially the sensor nodes), the hardware-, software-, and data-protection mechanisms must consider the possibility of physical access to the nodes.  

Research on securing the sensed data and the sensing devices has been mainly focused on the integration of trusted platform modules (TPM) and other secure cryptoprocessors into the sensors or host devices. The anonymous attestation feature of TPM is used to attest to the integrity and authenticity of the sensed data closer to the data source. Furthermore, a TPM attests the system state before sensitive information is transmitted.

Early work on securing sensor nodes~\cite{dua2009towards} was motivated by participatory sensing. The work made a case for trustworthiness in participatory sensing by content protection.  
Incorporation of TPM into mobile devices, participatory sensing application servers, and end user devices were proposed. The TPM attests the integrity of the sensed data in the mobile devices for submission to a participatory sensing application server. The proof of concept comprised an add-on circuit board, housing a TPM (TCG v1.2) chip, interfaced to a Nokia N800 phone. Overhead incurred due to the proposed solution amounted to $13$ kilobytes of memory (attestation code size), a latency of $1.92$ s (attestation time), verification latency of $0.78$ s.

Saroiu and Wolman~\cite{saroiu2010sensor} introduced the concept of trusted sensors and proposed the integration of a TPM functionality into  mobile device sensors to ensure integrity of the sensed data within the sensors. The work identified the IoT applications that would benefit from the deployment of trusted sensors. These included participatory sensing, monitoring energy consumption, and documenting evidence of crime scenes. A high-level conceptual design of a trusted sensor, in which a TPM was incorporated into a sensor, was presented. However, the work did not provide any proof of the concept. 

Dietrich and Winter~\cite{dietrich2009implementation} explored software TPM implementations for embedded systems. Existing CPU extensions like ARM TrustZone were evaluated to implement a software TPM with security guarantees similar to those of dedicated hardware. Aaraj et al.~\cite{aaraj2008analysis} also explored a software TPM solution. In order to achieve a performance improvement, critical functions were implemented on reconfigurable hardware.

Our earlier work, TrustCAM~\cite{winkler2010trustcam} and TrustEYE~\cite{winkler2014trusteye}, exploited TPM chips for protecting embedded camera nodes. TrustCAM used anonymous attestation and time-stamping features of the TPM to protect the integrity, authenticity and confidentiality of the image data on the host processor. To ensure image data integrity and authenticity, frame-groups were signed using a platform-bound key. Digital signing slowed down the frame rate only by $0.5$ frames per second compared to plain streaming. TrustEYE aimed to protect the captured images closer to the sensor. A TPM chip was integrated into the sensing unit, which has exclusive access to the sensor's data. Integrity, authenticity, confidentiality and freshness of the sensed data were protected at the sensing unit using 2048-bits RSA keys. A cartooning filter was implemented to preserve the privacy of monitored individuals \cite{erdelyi2014avss}. At a resolution of $320\times240$ a frame-rate of $9$ frames per second was achieved.

\begin{table}
\centering

%\resizebox{1.0\columnwidth}{!}{
\begin{tabular}{c c c c c}
\toprule
 & Approach & Data Security & Node Security & Usage Privacy \\
\toprule
%\rule{-2pt}{4ex}
~\cite{saroiu2010sensor} & TPM in sensor & yes & no & no \\
%\hline
~\cite{dua2009towards} & TPM in sensor & yes & no & no \\
%\hline
~\cite{winkler2010trustcam} & TPM in camera & yes & partial & partial\\
%\hline
~\cite{winkler2014trusteye} & TPM in image sensor & yes & no & no\\
%\hline
~\cite{potkonjak2010trusted} & PPUF in sensor & yes & no & no\\
~\cite{cornelius2008anonysense} & Mix networks & no & no & yes\\
~\cite{de2011short} & Sensed data encryption & yes & no & yes\\
~\cite{dimitriou2012pepper} & Token-based data access& no & no & no\\
\midrule
%\rule{-2pt}{4ex}
This work & Trusted sensing & yes & yes & yes\\
%\hline
This work & Secure node & yes & yes & yes\\
\midrule
\end{tabular}
%}
\caption{Classification of the related work on securing sensor nodes and contributions of this work}
\label{tab:sota}
\end{table}

Potkonjak et al.~\cite{potkonjak2010trusted} proposed a different approach for the trusted flow of sensed data in remote sensing scenarios. The approach employed public physically unclonable functions (PPUFs). PPUFs are fundamentally different from PUFs in several aspects: First, PPUFs are hardware security circuits which can be modeled by algorithms of high complexity whereas PUFs cannot be modelled. Second, the security of a PPUF relies on the fact that the PPUF hardware output is many orders faster than its software counterpart (i.e., model) whereas the security of a PUF relies on the unclonability of the PUF circuit. The major drawback of the PPUF-based approach lies in the fact that current PPUF designs involve complex circuits that require high measurement accuracy.  This slows down the authentication process and therefore the solution is not scalable. Additionally, the solution targets applications where privacy is not a concern. 

Some recent research efforts have lead to successful identification of PUF behavior on sensors. Sensor PUF is an idea introduced by Rosenfeld et al.~\cite{rosenfeld2010sensor} whereby the PUF response is determined by the applied challenge as well as the sensor reading. Cao et al.~\cite{cao2015cmos} introduced a CMOS image sensor based weak PUF. The PUF response bits are generated by comparing the random fixed pattern noise in selected pixel pairs. Although PUFs are lightweight, hardware security primitives that can be used to offer a scalable solution, identification of PUF behavior on sensors is only a part of the solution. 

Early work on usage privacy of personal sensing devices was also motivated by participatory sensing. Anonysense~\cite{cornelius2008anonysense} is a participatory sensing model that uses a trusted authority to anonymize the sensed data. Instead of submitting the sensed data directly to the application server the mobile devices submit the data to the anonymizing authority. The authority collects the sensed data from the participating mobile devices, anonymizes it, and forwards it to the application server.  Mobile devices communicate with the authority via Mix network. The application server assigns sensing tasks to the mobile devices using Tor anonymizing network. Anonysense offers $k$-anonymity, where $k$ is given by the number of mobile devices contributing sensed data  to the application server.  Anonysense had a number of limitations: First, observe that in order to guarantee $k$-anonymity, a Mix network may wait to receive $k$ reports before forwarding
them to the application server. This may significantly affect the service offered by the application.  Second, anonymization is performed after the data leaves the smartphone, whereas previously bugs~\cite{iPhonetracking} have successfully exploited the vulnerabilities in the software stack of the smartphone to leak users' privacy, thereby rendering the entire anonymization process ineffective. 
 
PEPSI~\cite{de2011short}, another participatory sensing framework, used identity based encryption for end-to-end encryption of sensed data reports. Smartphones register with  a trusted registration authority and obtain IDs corresponding to the application they intend to participate in. The application server only receives encrypted reports and forwards them to the intended end-user by matching the tags. The solution is only suitable for decentralized applications as the server cannot process the encrypted reports.

PEPPeR~\cite{dimitriou2012pepper} proposed a protocol for privacy-aware access of the sensed data by the end users (sensed data consumers) in participatory
sensing networks. An end user obtains tokens from the application server which reveal nothing about either the identity or its desire to spend the token with a
specific sensed data. The token validity, double-spending prevention are incorporated in the protocol using the classic cryptographic techniques.

An overview of the discussed related work is summarized in Table~\ref{tab:sota}. To summarize the previous work on sensed data trustworthiness, a TPM-based approach incurs significant hardware overhead on a node which may not be an economical solution for resource constrained sensor nodes. Despite widespread deployment of TPMs in laptops, desktops, and servers for over a decade, TPMs have not yet found their way into resource-constrained embedded devices. Moreover, TPMs do not provide protection against physical attacks. Due to open nature of IoT applications, sensors might be physically accessible to the attackers, which render TPM-based solutions ineffective in the given scenario. Protocols based on complex PPUF primitives are slow, have limited scalability and do not address the privacy protection. In this work, we identify PUF behavior on platforms that can serve as sensing platforms for a broad range of sensors. Furthermore, usage privacy of the sensors was not explicitly considered or addressed by any of the reviewed works.

In the related work on sensors' usage privacy, all proposed solutions are based on an online trusted authority. The online nature of the authority significantly increases the risk of keys compromise. Mix network based solutions are slow and may not be ideal for real-time or latency critical applications. Solutions leveraging end-to-end encryption of the sensed data are suitable only for the decentralized applications.   

By leveraging lightweight cryptographic techniques we propose effective solutions for protecting sensed data, sensor nodes and privacy of data producers, which are hooked into sensor hardware and are therefore harder to bypass. We present protocols for privacy-aware reporting of sensed data in IoT applications for both centralized and decentralized IoT applications. The solution does not uses an offline trusted authority which greatly reduces the risk of compromising keys. The protection of sensed data and privacy locally on the sensing devices, further reduces the risk of collusion and Sybil attacks.    

\section{Overall Approach for Trustworthy and Privacy-Aware Sensing} 
\label{overview}

IoT applications vary significantly in their infrastructure (e.g., cloud vs.~edge), sensed data collection mechanisms (e.g., raw data vs.~processed information collection), data processing requirements (e.g., processing at data source vs.~processing at server), data acceptance criteria and the services. A sensor-centric security solution to ensure sensed data trustworthiness and sensors' usage privacy depends on whether the processing of the sensed data takes place on the sensors or the server side. Therefore, with respect to data processing requirements, we categorize the IoT applications into two classes and propose two schemes tailored for the two classes of applications:

The first class of applications % , depicted in Fig.~\ref{fig:scenario1}, 
is attributed to the collection of raw data from the sensing devices. Processing of the data takes place at a central server. The sensors are (embedded or externally) connected to a host device that reads the sensors and relay the sensed data to the server. Participatory sensing applications are a common example  of centralized applications. 

We propose \textit{trusted sensing and privacy-aware reporting}  for the the centralized applications to ensure (i)  trustworthiness of  sensed  data (i.e., data with integrity, authenticity and freshness guarantees) and  (ii)  usage privacy (i.e., anonymity of  the sensing devices  and  unlinkability  of  multiple submissions from a device). The scheme works in two stages. 

First, the trustworthiness  of  sensed  data  is ensured  by \textit{trusted sensors}. Each trusted sensor extracts its unique fingerprint from the sensor hardware using on-chip physically unclonable functions (PUF) and attests to integrity and authenticity of each sensed reading by signing it using an identity-based signature scheme. The signature scheme uses the sensor-bound, unique fingerprint as the signing key. Second, to   report   the   signed   readings   from the  trusted  sensors in  a  privacy  preserving  manner, all  privacy leaking information (e.g., signature with a sensor bound unique key) is anonymized using a non-interactive   witness   indistinguishable   proof   system ($P_\mathrm{NIWI}$)~\cite{groth2012efficient}. Due  to  resource  constraints  on  the  sensors, we  offload  the  privacy  protection  mechanism  to  the host processor on the host CPU. Since  the  host  OS  is  assumed  to  be  untrusted,  we leverage a virtualization  approach~\cite{chen2008overshadow, brakensiek2008virtualization} where  the  user's software  environment  runs  as  a  guest  virtual  machine.  The root virtual machine, inaccessible to the user, has exclusive access to the trusted sensors and runs the privacy protection mechanism on the sensors' output. 

The second class of applications %, shown in Fig.~\ref{fig:scenario2},
leverages the resources on the host devices for processing the sensed data locally on the sensor nodes. Semantic information extracted from the data is delivered to the server. Visual monitoring applications are prominent examples that fall under this class of applications. 

With the \textit{trusted sensing and privacy-aware reporting} approach, once sensed data is signed at the sensor, any processing of the data at host devices invalidates  the security guarantees, which render the trusted sensing approach unsuitable for the given scenario. Instead, a holistic security solution encompassing the sensor and host device, called \textit{secure node}, is presented. The secure node approach addresses all layers of sensor node (sensor and host) stack including the applications, middle-ware, OS, and the hardware. % as illustrated in Fig.~\ref{fig:scenario2}.

We present the details of \textit{trusted sensing and privacy-aware reporting} and \textit{secure node} schemes for centralized and decentralized IoT applications in Sections~\ref{sec:TrustedSensing} and~\ref{securenode}, respectively.

\section{Trusted Sensing and Privacy-Aware Reporting}\label{sec:TrustedSensing}

\begin{figure}%[h]
\centering
\includegraphics[width=0.9\columnwidth]{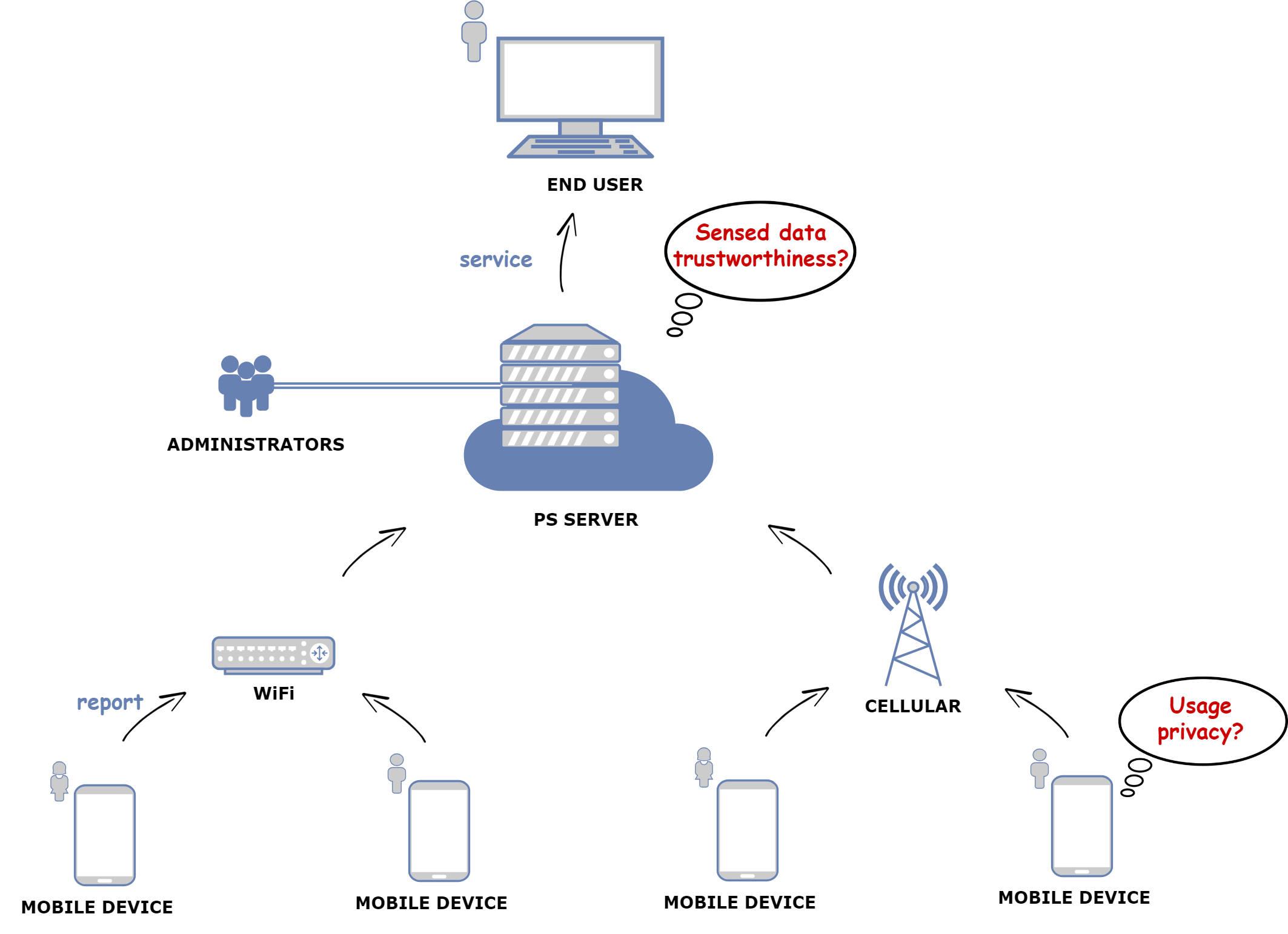} 
\caption{A high level infrastructure of participatory sensing (PS) applications}
\label{fig:PSI}
\end{figure}

This section presents the \textit{trusted sensing and privacy-aware reporting} approach for centralized applications. To illustrate our approach, we consider the participatory sensing (PS) scenario of Fig.~\ref{fig:PSI}.  The individuals interested in contributing sensed data to a PS application register their mobile devices with the PS server (also known as application server). During the registration,  a client software is downloaded and installed on the mobile device.  In order to contribute sensed data to the PS application, the client software running on the mobile device triggers a system call to the OS to read out the required sensors and return the readings to the client. The client composes them in form of a report and relays them to the server.  The mobile devices may use a WiFi or a cellular network Internet service to submit the sensed reports. The PS server collects reports from the contributing mobile devices, archives it for short- or long-term, and performs processing on the collected data. Processing includes filtering the high-quality data,  extracting information from the collected data and presenting the information in a format required by the end user. This information is provided to the end users as service.  

The trusted sensing and privacy-aware reporting aims  for  two  security objectives: (i) trustworthiness of  sensed  data  and  (ii)  anonymity of  sensing devices  and  unlinkability  of  multiple submissions from a device.

\begin{figure}%[h]
\centering
\includegraphics[width=0.8\columnwidth]{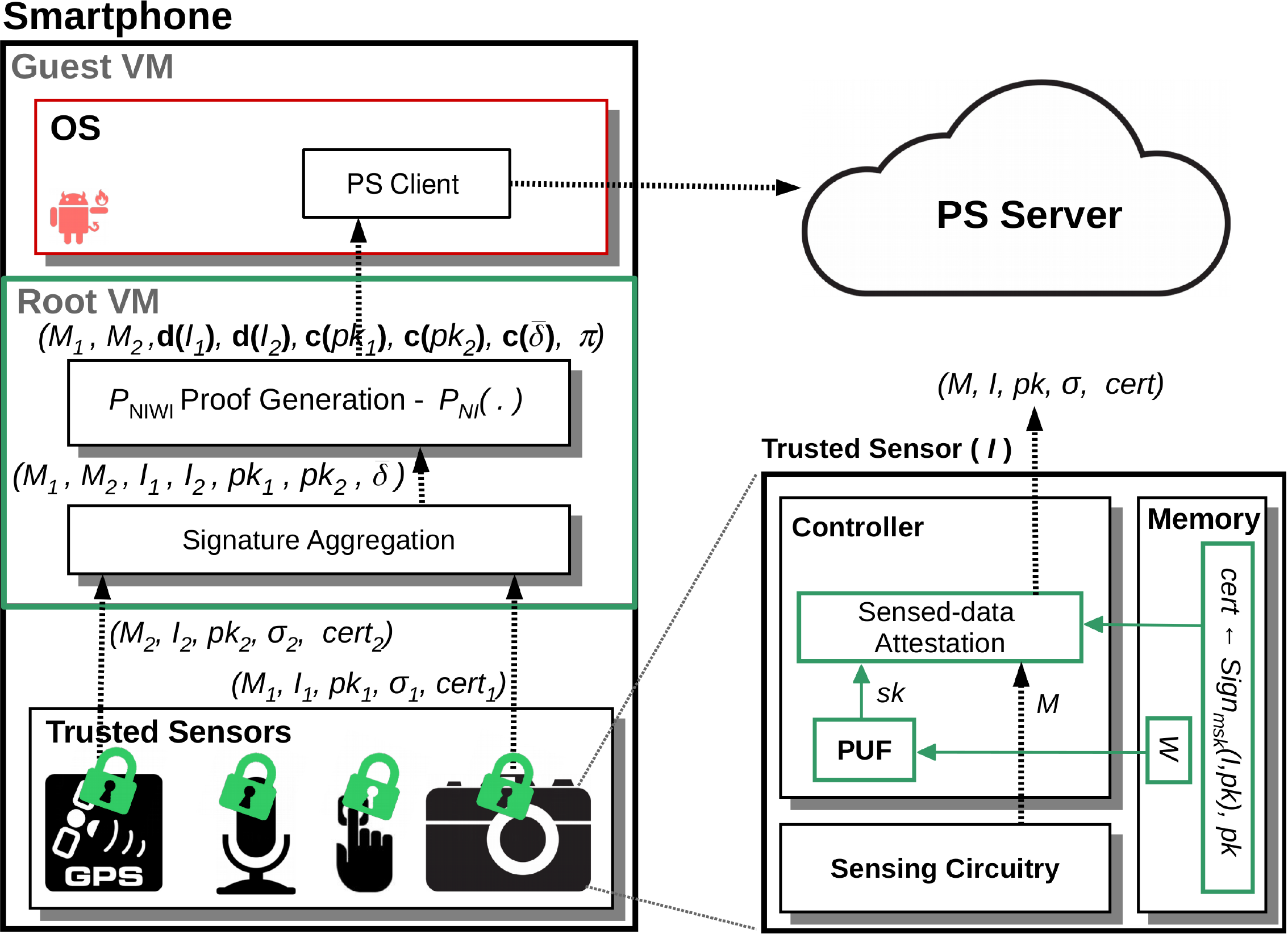}
\caption{Trusted sensing and privacy-aware reporting scheme: Trusted sensors provide integrity and authenticity guarantees on sensed data. Sensor readings are aggregated and   anonymized on the root virtual machine (VM) of the host device. The modules of the proposed security scheme are marked in green}
\label{fig: SolScn1}
\end{figure}

The trustworthiness  of  sensed  data  is ensured  by trusted sensors that comprise two key components: (i) a PUF framework extracts sensor fingerprint using an on-chip PUF and binds a unique key to the sensor hardware using the fingerprint and (ii) sensed data attestation uses an Identity-based Signature scheme to sign every sensor reading using the sensor-bound key (depicted  in  Fig.~\ref{fig: SolScn1}).  The scheme uses a trusted authority who securely binds a unique key to each sensor hardware.

To   report   the   signed   readings   from   trusted  sensors in  a  privacy  preserving  manner  we  adopt a Non-interactive   Witness   Indistinguishable   Proof   System ($P_\mathrm{NIWI}$)~\cite{groth2012efficient}.  Due  to  resource  constraints  on  the  sensors, we  offload  the  privacy  protection  mechanism  to  host processor on user device.  Since  the  host  OS  is  assumed  to  be  untrusted,  we leverage  virtualization  approach~\cite{chen2008overshadow, brakensiek2008virtualization}  where  the  user's software  environment  runs  as  a  guest  virtual  machine.  The root virtual machine is inaccessible to the user.

To ensure anonymity and unlinkability of multiple submissions by a user device, uniquely identifying information in a trusted sensor's output such as a sensor's signature on the reading using the unique sensor-bound key cannot be revealed to  the  server since  it  can  uniquely  identify  the sensor, thereby the user device and the user. Instead,  the  mobile device  computes and reports proof  of  knowledge  of the uniquely identifying information. This is done using $P_\mathrm{NIWI}$. 

% , i.e., identifier, sensor's signature on the reading, and  the  certificate from the authority
Given  a  mobile device embedded  with  the trusted sensors,  the  root  virtual  machine executes the prover algorithm of the $P_\mathrm{NIWI}$ as follows: (i) read  the  trusted  sensors,  (ii)  commit  to  the  witness (i.e., uniquely  identifying  information  that  we  want  to  anonymize such as sensor identity, the signature, and the certificate) and (iii)  generate  the  proofs  of  knowledge  of  the  sensor  identity,the  signature,  and  the  certificate.  These  commitments  and proofs are sent to the server along with the sensed readings.  The server executes the verifier algorithm of the $P_\mathrm{NIWI}$ using received readings,  commitments,  and  the  proofs as arguments and verifies that  the  prover in fact possesses  a valid  signature-certificate  pair  for  each  received reading, thereby verifying the integrity and authenticity of the readings.

The  witness-indistinguishability  of  the  $P_\mathrm{NIWI}$ proof  system implies that the commitments and proofs do not reveal (to the server) the witness  used  to construct the commitments and the proofs. Anonymity of the prover is given by the number of possible witnesses. Given $N$ mobile devices equipped with trusted sensors and submitting readings to the PS server, anonymity of each user is given by $N$.  Our  scheme  aggregates all signatures and certificates in the trusted sensors' output into a single signature and then generates the proof of knowledge of the aggregate signature, as  illustrated  in  Fig.~\ref{fig: SolScn1}. This considerably reduces the communication overhead incurred on the user device.

Next, we present the trusted sensing and privacy-aware reporting components of the security scheme in Sections~\ref{sec: ts} and~\ref{PrivReport}, respectively.

\subsection{Trusted Sensing} 
\label{sec: ts}

Trusted sensing is accomplished by the trusted sensors in two steps: fingerprint extraction and sensed data attestation. The former refers to extraction of unique, non-transferable fingerprints from the sensor hardware by a legitimate authority in a secure environment. The later uses a digital signature scheme to attest the integrity and authenticity of sensor readings. The sensor fingerprint serves as the signing key for the sensed data attestation. The receiver (e.g., PS server) can verify the integrity and authenticity of each reading by signature verification. The proposed trusted sensor uses a \textit{PUF framework} to extract the sensor fingerprint and binds it to the sensor hardware. An \textit{Identity Based Signature Scheme (PUF-based Cert-IBS)} is used for the sensed data attestation which ensures non-repudiation of the data. 

\subsubsection{PUF Framework}
\label{subsec:PUF_framework}

Physically unclonable functions (PUF) are special lightweight circuits that use the CMOS manufacturing process variations to generate the fingerprint of the underlying hardware. Typical attributes of a PUF include randomness, uniqueness, physical unclonability,  and reliability. A PUF circuit provides a challenge-response mapping that is based on the uncontrollable variations in the physical structure of the integrated circuit (IC) introduced during the manufacturing process. These variations are random and unique for each instance which makes any PUF-enabled electronic hardware uniquely identifiable (uniqueness). Moreover, the chip manufacturer is not able to control or forge these variations (physical unclonability).  Reliability implies that a PUF should be able to reproduce the same challenge-response pairs under a range of environmental and operating conditions. However, in practice, multiple responses from a PUF instance obtained under different environmental conditions (e.g., temperature) or operating conditions (e.g., voltage supply) slightly differ from one another. These variations are referred to as PUF noise or error-rate and are measured as intra-Hamming distance ($HD^{intra}$). Uniqueness of a PUF mapping is measured in terms of inter-Hamming distance ($HD^{inter}$) which is a measure of how  different  two  responses  from  two  PUF  instances  are. Randomness of a PUF response is measured in terms of Hamming weight ($HW$) of the response. Ideally, maximum $HD^{intra} \approx 0 \% $, average $HD^{inter} \approx 50 \% $ and average $HW \approx 50 \% $.

In order to extract an uniformly distributed random and perfectly reproducible fingerprint from the noisy and biased PUF response, helper data algorithms (HDAs) are used. Our scheme requires the flexibility of masking an externally generated cryptographic key with the device fingerprint; therefore we use the HDA by Tuyls~\cite{tuyls2006rfid}. The PUF framework is comprised of two modules: the PUF and the HDA and works in two phases: key binding and key extraction.

\begin{enumerate}    

\item Key Binding: $W \leftarrow Gen(r, k)$\\ It is a one-time protocol carried out by a legitimate authority on the PUF in a secure environment to generate helper data $W$. A challenge $c$ is applied to the PUF and a response $r$ is obtained. The authority then chooses a random key $k\in\{0,1\}^k$ and calculates the corresponding helper-data as $W\gets r\mathbin{\oplus}C_\mathrm{k}$, where $C_\mathrm{k}$ is the nearest code-word chosen from the error-correcting code $\mathcal{C}$, with $2^k-1$ code-words. $W$ is integrity protected public information.  

\item Key Extraction: $k \leftarrow Rep(r',W)$\\ It is performed every time the key extraction from the PUF is desired. The PUF is subjected to the same challenge $c$ and a noisy response $r'$ is obtained. The code-word is then calculated as $C_\mathrm{k'} \leftarrow r' \oplus W$. If $r'$ corresponds to the same challenge $c$ applied to the same PUF, $k$ is obtained after decoding $C_\mathrm{k'}$ using $W$ otherwise an invalid code-word is obtained i.e., $k \leftarrow \mathtt{Decoding}(C_\mathrm{k'}),$ if $\ \mathtt{Hamming\ distance}(C_\mathrm{k},C_\mathrm{k'}) \leq t$, where $t$ is error-correction capacity of $\mathcal{C}$.
%Note that to generate the key $k$, the sensor has to perform only an XOR and a decoding operation. 
\end{enumerate}

This PUF framework offers the following key advantages: (i) it binds a unique key with a PUF-enabled hardware, (ii) it provides secure storage of the key since the key is derived from device properties during start-up and (iii) it offers more cost-effective secure key storage than a secure memory alternative. 

\subsubsection{Identity Based Signature Scheme (PUF-based Cert-IBS)}
\label{PUFIBS}

This section explains our Identity based Signature scheme (PUF-based Cert-IBS) that is based on the framework~\cite{bellare2009security} to construct certificate-based identity-based signature scheme (Cert-IBS) from a standard signature (SS) scheme. PUF-based Cert-IBS ensures integrity and authenticity of sensed data in our \textit{trusted sensing and privacy-aware reporting} and \textit{secure node} schemes. It uses the PUF framework of Section~\ref{subsec:PUF_framework} to bind the security key with the sensor fingerprint extracted using PUF. A typical SS comprises three algorithms: key generation $(K)$, signing $(Sign)$ and verification $(Ver)$. PUF-based Cert-IBS uses a key generation authority. To setup PUF-based Cert-IBS, the authority generates a master key pair $(msk, mpk)$ using $K$. 

We assign an identity $I$ and a PUF instance $PUF$ to each sensor. The identity can be any unique physical identifier of the sensor such as serial number, EPC or a unique bit string written to one-time programmable memory of the sensor. We denote a sensor with identity $I$ and $PUF$ by SEN($I,PUF$). 

\begin{enumerate}

\item \textbf{Setup.} The trusted authority runs the $K$ of $SS$ to generate the master key pair: $(mpk, msk)\gets K(1^k)$

\item \textbf{Enrollment.} During the enrollment phase, the authority generates a unique signing key pair $(sk, pk)$ using the key generation algorithm $K$ of SS and binds $sk$ with the on-chip $PUF$ using the key binding algorithm of the PUF framework, i.e., $W_{sk}\gets Gen(r, sk)$  where $r$ is the $PUF$ response to challenge $c$ selected by the authority and $W_{sk}$ is the helper data corresponding to $sk$. Further, the authority issues a certificate on the public half of the signing key given by $cert \gets Sign_{msk}(pk, I)$. $W_{sk}$ and $cert$ are stored in the sensor's non-volatile memory. 

\item \textbf{Sensed Data Attestation.} %We refer to PUF-based Cert-IBS signing as sensed data attestation since by signing the sensor attests to integrity and authenticity of the sensed reading. 
Sensed data attestation is performed every time the sensor SEN($I, PUF$) outputs a new reading. The private key required for signing is reconstructed at the power-up using the key extraction phase of the PUF  framework i.e., $sk\gets Rep(r',W_{sk})$. PUF-based Cert-IBS signature of SEN($I, PUF$) on sensor reading $M$ is given by $(M, I, pk, \sigma, cert)$, where $\sigma \gets Sign_{sk}(M)$. PUF-based Cert-IBS verification is successful if $Ver_{pk}(M, \sigma) = 1$ and $Ver_{mpk}((I,pk), cert) = 1$. Successful Cert-IBS verification ensures that reading $M$ is signed by SEN($I, PUF$) with its platform-bound private key, assigned and bound to SEN($I,PUF$) by the legitimate authority. 
\end{enumerate}

Given that $SS$ is a uf-cma secure standard signature scheme, theorem 3.5 of~\cite{bellare2009security} proves that the corresponding $\mathrm{PUF}$-$\mathrm{based\ Cert}$-$\mathrm{IBS}$ as per construction of Section \ref{PUFIBS} is a uf-cma secure IBS scheme.

\subsection{Privacy-Aware Reporting of Trustworthy Sensed Data}
\label{PrivReport}

Privacy-aware reporting of sensed data entails anonymity of user devices and unlinkability of multiple submissions from a user device. Given a user device incorporated with the trusted sensors, each element of tuple $(I, pk, \sigma, cert)$ in a trusted sensor's output uniquely identifies the sensor and therefore cannot be revealed to the data center. For each trusted sensor output $(M, I, pk, \sigma, cert)$, the user device computes a proof of knowledge of $(I, pk, \sigma, cert)$ using the non-interactive witness indistinguishable proof system ($P_\mathrm{NIWI}$) by Groth and Sahai ~\cite{groth2012efficient}. The mobile device (prover) then sends the proof instead of $(I, pk, \sigma_{M}, cert)$ along with the sensor reading $M$ to the server (verifier) as depicted in Fig.~\ref{fig: SolScn1}. The server verifies the proof. Successful verification ensures that the mobile device knows a witness $(I, pk, \sigma, cert)$ such that PUF-based Cert-IBS verification equations holds true for the received data $M$ i.e., $Ver_{pk}(M,\sigma)=Ver_{mpk}((I,pk), cert)=1$. 
 
Given $N$ trusted sensors submitting sensed data to a server,  witness indistinguishability of the proof implies that the data center cannot distinguish which witness $\{(I_i, \sigma_{i},cert_i)\}_{i=1}^N$ (i.e., trusted sensor) was used to construct the proof. Therefore, every trusted sensor is $N$-anonymous with respect to the server. Unlinkability of multiple submissions by the same sensor follows from witness indistinguishability. 

\subsubsection{Primer on Pairings}

For privacy-aware reporting of sensed data, our scheme uses pairings-based cryptography, therefore we sketch some basics of pairings: Let $G_1$, $G_2$ and $G_T$ be cyclic groups of the same prime order and $g_1$ and $g_2$ are the generators of $G_1$ and $G_2$ respectively.  A pairing is map $e:G_1\times G_2 \rightarrow G_T$ that is (i) bilinear, i.e., for all $u\in G_1$, $v \in G_2$ and $a$,$b\in \mathbb{Z}, e(u^a,v^b) =e(u,v)^{ab}$, (ii) $e(g_1,g_2)$ generates $G_T$ and (iii) $e$ is efficiently computable.
The setting where $G_1 = G_2 = G$ and $g_1 = g_2 = g$ is called symmetric pairing whereas if $G_1 \neq G_2$ and $g_1 \neq g_2$, the pairing is called asymmetric. For simplicity, we explain our scheme for a symmetric setting. For practical implementation, an asymmetric setting is recommended (cp.~Section~\ref{eff}).

\subsubsection{Non-interactive Witness Indistinguishable Proofs $(P_\mathrm{NIWI})$}\label{NIWI}

A proof system allows a prover who possesses some witness $\omega$ to convince a verifier that a certain statement $\chi \in L$ is true, where $L$ is some language, and $\omega$ is a witness that attests to this fact. In witness indistinguishable proof systems, the interaction between the prover and the verifier does not reveal information about the witness, even if the verifier behaves maliciously. Furthermore, it is unfeasible for an adversary to decide which of the possible witnesses is used by the prover. In a non-interactive proof system, the prover simply sends the verifier a single message after which the latter verifies correctness of the proof without any further interaction with the prover.

Groth and Sahai  introduced a non-interactive witness indistinguishable proof system ($P_\mathrm{NIWI}$) for languages involving the satisfiability of equations over bilinear groups, in the common reference string model~\cite{groth2012efficient}. The main idea underlying $P_\mathrm{NIWI}$ is as follows: Given groups $A_1, A_2, A_T$ with a bilinear map, $P_\mathrm{NIWI}$ maps the elements in $A_1, A_2, A_T$ into $B_1, B_2, B_T$, also equipped with a bilinear map, by using a commitment scheme. The latter groups are larger thereby allowing to hide the elements of $A_1, A_2, A_T$.

Given the equation(s) that we intend to prove, we replace the variables (witness) in the equation(s) with commitments to those variables. Since the commitments are hiding, the equations will no longer be valid. However, we can extract out the additional terms introduced by the randomness of the commitments and provide these terms in the proof to the verifier, who can verify the validity of the equations. Does providing these terms destroy witness indistinguishability? Since there are multiple additional terms introduced by substituting the commitments, the algebraic environment allows us to randomize the terms such that their distribution is uniform over all possible terms satisfying the equations.

By definition, $P_\mathrm{NIWI}$ is a tuple of four probabilistic  polynomial  time algorithms $(K_{NI}$, $P_{NI}$, $V_{NI}$, $X_{NI})$, i.e., key generator, prover, verifier, and extractor, respectively.  The key generator, $K_{NI}$, takes the bilinear group description $A_1, A_2, A_T$ as input and outputs a common reference string $crs$ and an extraction key $xk$. $crs$ comprises the target groups description $B_1, B_2, B_T$ and rules to compute commitments. Given a set of equations (that the prover wants to prove), the prover, $P_{NI}$, takes $crs$ and a witness $\omega$ as input and outputs a proof $\pi$ for each equation. The verifier, $V_{NI}$, given $crs$, the set of equations and $\pi$ outputs $1$ if the proof is valid and $0$ otherwise. Finally, the extractor, $X_{NI}$, on a valid proof $\pi$ may extract $\omega$ using the extraction key $xk$.

The privacy-aware trusted sensing scheme adopts $P_\mathrm{NIWI}$ for privacy-aware reporting of sensed data from the trusted sensors. During the setup, the TA runs the key generator algorithm $K_{NI}$, which takes the group description $(G,G_T, g, e, \mathrm{p})$ as input and outputs the $csr$ that comprises eight group elements, i.e., $\in G^8$ and an extraction key $xk$. The $csr$ is published for the participating mobile devices and the application servers where as $xk$ is kept secret by the TA since the extraction of the witnesses is not required in the proposed privacy-aware trusted sensing scheme. The scheme, however, can be extended with dispute resolution and revocation mechanisms using the extractor algorithm and the $xk$. A mobile device runs $P_{NI}$ every time it has to submit sensed data using the trusted sensors to an application server. 

Without a privacy-protection mechanism, a mobile device submits the trusted sensor reading $( M, I, pk,\sigma, cert)$ as is to the server who verifies the signature and the certificate using the PUF-based Cert-IBS verification equations, 	$Ver_{mpk}(I,cert)\myeq 1$ and $Ver_{pk}(M, \sigma)\myeq 1$. In this case, the server requires $(I, pk,\sigma, cert)$ as input to the PUF-based Cert-IBS verification algorithm. Each element of this tuple uniquely identifies the trusted sensor and the mobile device. With the $P_\mathrm{NIWI}$, the root virtual machine running on the mobile device commits to each element of  the witness $(I, pk,\sigma, cert)$ using the $csr$, $\mathbf{d}(I),\  \mathbf{c}(pk),\ \mathbf{c}(\sigma)$, and $\mathbf{c}(cert)$, where  $\mathbf{c}(.)$ denotes a commitment to an element in group $G$ whereas $\mathbf{d}(.)$ denotes a commitment to an element in $\mathbb{Z}_p$. The $pk,\sigma,\text{ and } cert \in G$ each. Although, $I$ is only assumed to be a unique, random value, one can use $I\mod p \ (\in \mathbb{Z}_p)$ instead.  Each  $\mathbf{c}(.)$ and $\mathbf{d}(.)$ is $\in G^3$. It further computes two proofs $\pi_1$ and $\pi_2$ by replacing the witness with the commitments in the PUF-based Cert-IBS verification equations. Since we use the symmetric version of the BLS signature scheme~\cite{boneh2004short} as the $SS$ in PUF-based Cert-IBS, the PUF-based Cert-IBS verification equations, denoted as $eq_1$ and $eq_2$, are given by Eqs.~\ref{eq1a} and~\ref{eq1b}. Each of the $\pi_1$ and $\pi_2$ $\in G^9$. The proofs and commitments are are then sent to the server along with the sensor reading $M$. The PS server runs the verification algorithm $V_{NI}$ to verify whether or not the proofs $\pi_1$ and $\pi_2$ and the commitments satisfy the following equations

\begin{subequations}\label{eq1}
\begin{eqnarray}
\label{eq1a}
&e(h_{M}, pk)= e(\sigma, g) \\ 
\label{eq1b}
&e(h_{c}, mpk)=e(cert, g)
\end{eqnarray}
\end{subequations}
\noindent
where $H(I, pk)$ is denoted by $h_{c}$ and $H(M)$ is denoted by $h_{M}$.

In PS applications a sensed data report comprises typically multiple sensors' readings. For a report comprising $Q$ sensors' readings, the above process is performed $Q$ times. The root virtual machine then provides the tuple $\{M_i,\ \mathbf{d}(I_i),\mathbf{c}(pk_i),\mathbf{c}(\sigma_{i}),\mathbf{c}(cert_i),\  \pi_{1i},\ \pi_{2i}\}_{i=1}^Q$ to the application client running in guest virtual machine on the user device, which sends it to the PS server, who runs the verification algorithm $V_{NI}$ for each sensor reading. For mathematical details and security proofs of the $P_\mathrm{NIWI}$ construction, the reader is referred to~\cite{groth2012efficient}.  

Theorem 17 in~\cite{groth2012efficient} proves that $P_\mathrm{NIWI}$ following the construction of Section~\ref{NIWI} has perfect completeness, perfect soundness and composable witness indistinguishability for satisfiability of Eqs.~\ref{eq1a} and~\ref{eq1b} in a bilinear group $G$ where DLIN problem is hard.  Witness indistinguishability implies that the proofs and the commitments do not reveal what values of the witness $(I,pk, \sigma, cert)$ were used to generate the commitments and the proofs. Anonymity of the prover $P_{NI}$ with respect to $V_{NI}$ is given by the possible number of values a witnesses can take. Given the PS scenario, anonymity of each trusted sensor with respect to the application server is given by the total number of the trusted sensors contributing sensed data to the server.

Assuming the DLIN assumption holds in $G$, the $P_\mathrm{NIWI}$ for each sensor reading costs $30$ elements of group $G$, i.e., four commitments $\mathbf{d}(I),\ \mathbf{c}(pk),\ \mathbf{c}(\sigma),\  \mathbf{c}(cert)$  consisting of $3$ group elements each and two proofs $\pi_1$ and $\pi_2$ of $9$ group elements each. A report comprised of $Q$ sensors' readings therefore incurs a communication overhead of $30Q$ elements of $G$ on the mobile device. 

\subsubsection{Communication Overhead Reduction using Signature Aggregation}

The aggregation property of BLS signatures allows an aggregating party to combine multiple, say $m$, signatures into a single signature as $\bar{\sigma} \gets \Pi_i^m\sigma_i$, where $\bar{\sigma} \in G$, thereby reducing the total signatures' size to $1/m$. For aggregate verification, given $\bar{\sigma}$, the original messages $M_i$, and public keys $pk_i$, compute $h_{M_i} \gets H(M_i)$ and accept if $e(\bar{\sigma},g)=\Pi_i^m e(h_{M_i},pk_i)$. %The BLS aggregate signature scheme is secure against existential  forgery in the aggregate chosen key model~\cite{boneh2003aggregate}. 
In our scheme, aggregation is done by the root virtual machine on the mobile device as follows: In a PUF-based trusted sensor output $\sigma_{i}$, i.e., $Sign_{sk_i}(M_i)$, and $cert_i$, i.e., $Sign_{msk}(I_i,pk_i)$, are both BLS signatures and can be aggregated. Furthermore, if an application requires every mobile device to submit multiple, say $Q$ sensor readings, the aggregation is done as $\bar{\sigma} = \Pi_{i=1}^Q \sigma_i \cdot cert_i$. 

A major reduction in communication overhead is achieved since $P_\mathrm{NIWI}$ is applied on the aggregate of $Q$ readings instead of individual ones. The simultaneous satisfiability of PUF-based Cert-IBS aggregate signature verification is given by Eq.~\ref{aggVer}. The root virtual machine sets the witness to $(\{I_i, pk_i\}_{i=1}^Q,\bar{\sigma})$ and equation we want to verify, denoted as $eq$, to Eq.~\ref{aggVer}. The prover, using $P_{NI}$, commits to each element of the witness and generates a proof $\pi$ by plugging the commitments to the $eq$. Successful verification using $V_{NI}$ at the server ensures that the prover (mobile device) possesses $Q$ PUF-based Cert-IBS signatures on $Q$ distinct readings $\{M_i\}_{i=1}^Q$ such that:

\begin{equation}\label{aggVer}
e(\bar{\sigma}, g)= \Pi_{i=1}^Q e(h_{M_i}, pk_i)e(h_{c_i}, mpk)
\end{equation}
 
The privacy-aware trusted sensing scheme is summarized in Table~\ref{FullScheme} and runs in three phases: setup, enrollment and trusted sensing and privacy-aware reporting. The TA sets up the scheme by generating its master key pair. The bilinear group description and $csr$ are also generated and published for the PS entities during the setup. Enrollment of the privacy-aware trusted sensing scheme is in fact the enrollment of the PUF-based Cert IBS enrollment, performed only once in a secure and trusted environment. The trusted sensing and privacy-aware reporting is performed every time a mobile device contributes sensed data to a PS application: The trusted sensors' output readings with integrity and authenticity guarantees, which are anonymized by the root VM on the host mobile device and sent to the PS server. The PS server verifies the integrity and authenticity of the readings in a privacy-preserving manner using the $P_\mathrm{NIWI}$ verification algorithm, $V_{NI}$.

$P_\mathrm{NIWI}$ of $Q$ aggregated readings for satisfiability of $eq$ costs $(6Q+12)$ elements of $G$ compared to $30Q$ elements of $G$ without aggregation. Anonymity of the mobile device with respect to the PS server is given by the total number of mobile devices reporting the PUF-based trusted sensors' readings to the PS server. 

\begin{table*}

%\begin{adjustbox}{angle=90}
\resizebox{\linewidth}{!}{
	\begin{tabular}{l}
	\toprule
    \rule{0pt}{4ex} 
    $\mathbf{Setup:}$ \quad TA\\
    \quad TA: \quad$H, \Sigma \gets \mathcal{G}(1^k) \mathrm{\ where}\ \Sigma= (G,G_T, g, e, \mathrm{p}) \ and\ H= \mathrm{collision\ resistant\ hash\ function}$\\
	\quad TA: \quad$(mpk,msk)\gets K(1^k)$\\
    \quad TA: \quad$(csr, xk) \gets K_{NI}(\Sigma)$\\
    \midrule
    \rule{0pt}{4ex} 
    $\mathbf{Enrollment:}\quad$SEN$(I, PUF) \leftrightarrow$ TA$(mpk, msk)$\\
     %\quad$usk \gets ENROL_{\mathrm{IBS}}(I, PUF)$\\
     \quad TA: \quad$W\gets Gen(r, sk)$\\
     \quad TA: \quad$cert \gets (pk, Sign_{msk}(pk,I))$\\
     %\quad$h_{FW} \gets ENROL_{\mathrm{VB}}(PUF, FW)$\\
     \midrule
     \rule{0pt}{4ex} 
     $\mathbf{Trusted\ Sensing\ and \ Privacy}$-$\mathbf{Aware\ Reporting:}\quad$SEN$(I, PUF, W, cert) \leftrightarrow$ MOB($\Sigma, H, csr$)$\ \leftrightarrow$ PS SERVER($\Sigma, H, csr$)\\
     \quad SEN$(I, PUF, W, cert)$:\quad$(M,I,\sigma, cert) \gets SDA(M)$\\
     \quad MOB (Root VM):\quad$(\{M_i, I_i,  pk_i\}_{i=1}^Q,\bar{\sigma}) \gets AGG(\{M_i, I_i,\sigma_i, cert_i\}_{i=1}^Q)$, for $Q$ sensor readings\\
     \quad MOB (Root VM):\quad$(\{\mathbf{d}(I_i), \mathbf{c}(pk_i)\}_{i=1}^Q,\ \mathbf{c}(\bar{\sigma}),\ \pi)\gets P_{NI}(\Sigma, csr,(\{I_i , pk_i\}_{i=1}^Q, \bar{\sigma}), eq)$\\
     \quad PS SERVER:\quad$1 \myeq V_{NI}(\Sigma, crs,\{\mathbf{d}(I_i), \mathbf{c}(pk_i), H(M_i)\}_{i=1}^Q,\ \mathbf{c}(\bar{\sigma}),\ \pi)$\\
    \midrule
    \rule{-3pt}{4ex} 
    $\mathrm{\mathbf{Notations:}}$ TA=Trusted Authority, MOB= Mobile device, $SDA$= Sensed data attestation, $AGG$= Signature aggregation
	\end{tabular} }
\caption[Privacy-aware trusted sensing scheme]{The Privacy-aware trusted sensing scheme is executed in three phases: Setup, enrollment and trusted sensing and privacy-aware reporting. The TA sets up the scheme by generating its master key pair. The bilinear group description and $csr$ are also generated and published for the PS entities. During enrollment, each sensor is enrolled with the TA using PUF-based Cert IBS enrollment. During the trusted sensing and privacy-aware reporting phase, the trusted sensors output readings with integrity and authenticity guarantees, which are anonymized by the root VM on the host mobile device and sent to the PS server. The PS server verifies the integrity and authenticity of the readings in a privacy-preserving manner using the $P_\mathrm{NIWI}$ verification algorithm $V_{NI}$}
\label{FullScheme}
%\end{adjustbox}
\end{table*}

\section{Secure Node}\label{securenode}

This section presents the \textit{secure node} approach for decentralized IoT applications. 
Visual sensor networks (VSNs) are becoming increasingly popular in IoT applications that range from surveillance of critical public spaces for law and order maintenance and public safety to private space monitoring such as smart  homes,  assisted/enhanced living,  child  monitoring  and  home security~\cite{Winkler_COMPSURV2014}.  Continuous transmission of visual data requires high bandwidth and memory which is often unfeasible.  Therefore, visual monitoring applications typically require processing of visual data locally on camera nodes. A major limitation of the trusted sensing and privacy aware reporting approach of Section~\ref{sec:TrustedSensing} is that once data is signed within the sensor, any legitimate modification (processing, compression etc.) of the data at host processor invalidates the security guarantees. With secure node approach, we overcome these limitations by adopting a holistic security solution for the node addressing all layers of a camera stack including the applications, middle-ware, OS, and the hardware.

\begin{figure}
\centering
  \includegraphics[width=\linewidth]{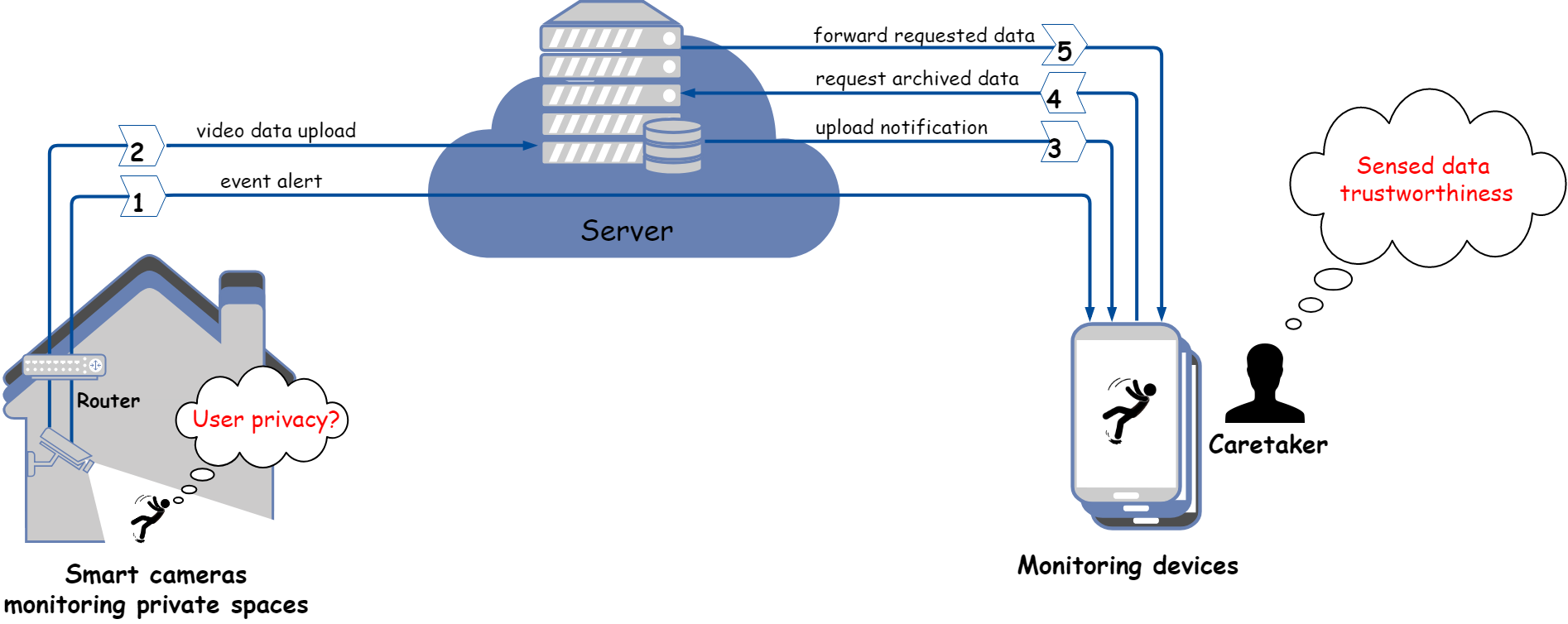}
  \caption{A high level infrastructure of private space monitoring applications depicting security and privacy requirements}
  \label{fig:sys}
\end{figure}

To illustrate our approach, we consider a visual monitoring for assisted living scenario of Fig.~\ref{fig:sys} where one or more cameras monitor the space for events of interest  such as fall detection or no movement for long duration. To limit the amount of data transmitted by the camera device, archiving is triggered upon event detection. The event can be triggered internally (e.g., by on board analytics) or externally (e.g., by auxiliary sensors or a request from the caretaker). We assume that integrity and authenticity of an external source is verified before triggering an event. Upon event detection, an alert message is sent to a caretaker and the video data capturing the event is uploaded to a storage server. When the upload has been completed, a push  notification  is  sent by  the  server to the caretaker who then downloads the data, analysis it and formulates a response.

In order to keep the cost of the camera low, we do not assume availability of permanent storage of videos on the camera node. A public cloud storage server is leveraged for short or long-term video archiving. To limit the amount of data transmitted by the camera device, archiving is triggered upon event detection. The event can be triggered internally (e.g., by onboard analytics) or externally (e.g., by auxiliary sensors or a request from the end user). We assume that integrity and authenticity of an external source is verified before triggering an event. 

Data security, node security and personal privacy of the monitored individuals are the integral parts of the sensing device. Data security includes integrity, authenticity, confidentiality, and freshness of the visual data and the metadata. Data security is ensured on the sensing device before it is delivered to the server. Since visual data contains identities and behaviors of observed individuals, data confidentiality and access authorization are essential security requirements for personal privacy protection in visual monitoring applications. These security guarantees are valid throughout the entire lifetime of the data.

 Confidentiality is ensured by encrypting each video frame using AES128 encryption. Integrity and authenticity are ensured by signing the hash-chain of encrypted frames  using the PUF-based Cert-IBS scheme of Section~\ref{PUFIBS}. PUF-based Cert-IBS and AES128 algorithms use platform-bound security keys. PUF framework of Section~\ref{subsec:PUF_framework} binds the signing and encryption keys to the camera hardware using on-chip PUF that serves as secure key storage. On event detection, encrypted-hashed-signed footage is uploaded to a public storage server at the edge or cloud tier and an alert message is sent to the end-user who can then download the archived footage from the server on demand. Integrity, authenticity and freshness of data is ensured by verifying PUF-based Cert-IBS signatures. Only the authorized end-user (i.e., having access to the decryption key) can decrypt the frames. Since data security is implemented at application level, in order to ensure effective security guarantees, underlying software and hardware stack of the camera node needs to be protected as well. Node security requirements include integrity, authenticity and unclonability of camera firmware, resistance against hardware tampering and side-channel attacks.

\begin{figure}
\centering
  \includegraphics[width=0.9\linewidth]{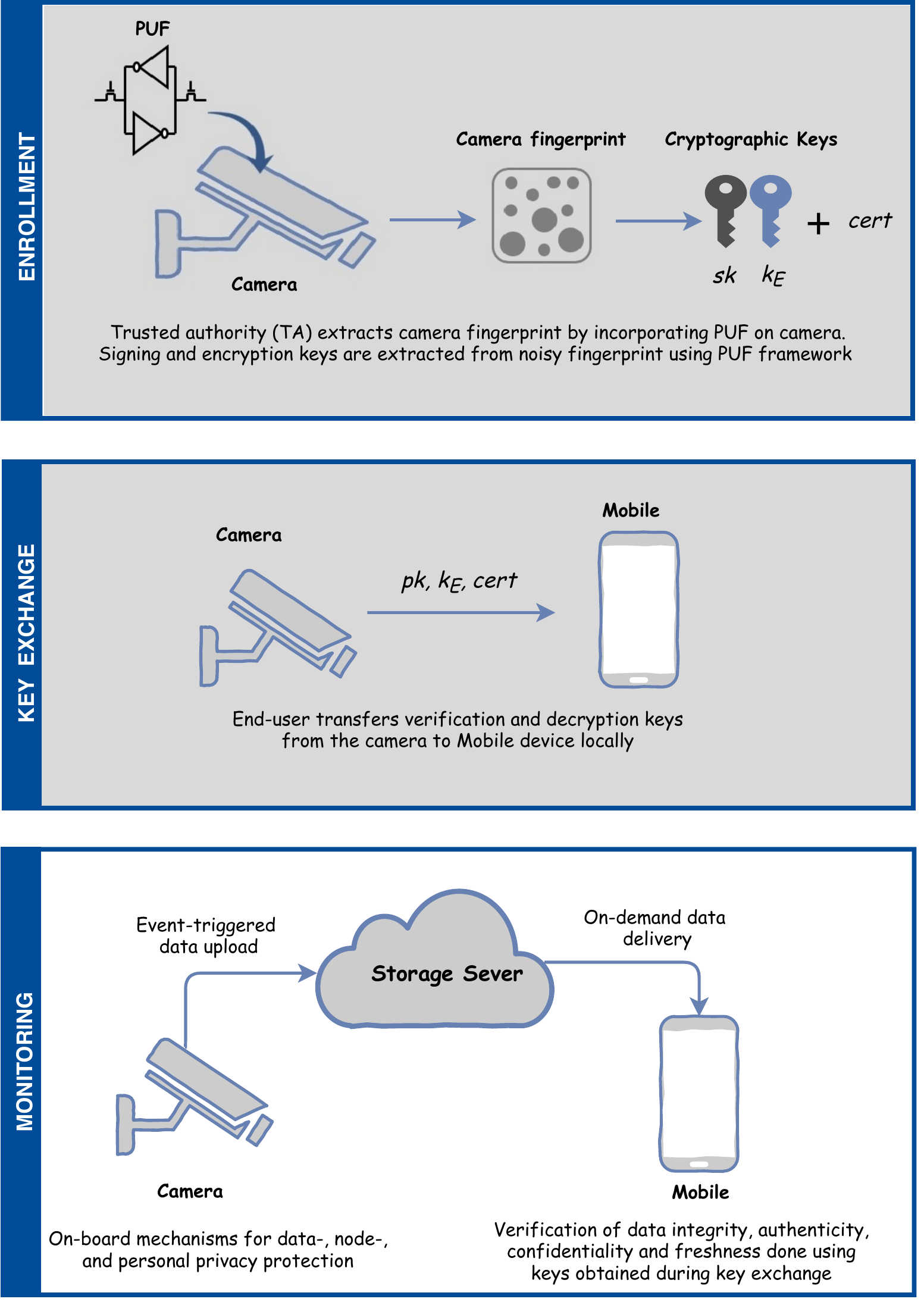}
  \caption{The enrollment, key exchange and monitoring phases of the secure node scheme for visual monitoring for assisted living. The enrollment is performed once in a secure and trusted environment whereby the platform-bound unique cryptographic keys are created. During key exchange, the end user shares verification keys with caretaker's monitoring device using a local interface. Thereafter, the camera is deployed for monitoring}
  \label{fig:Phases}
\end{figure}

\subsection{Operational Phases}
\label{OP}

In order to bind the cryptographic keys with the camera platform and limit data access to only legitimate caretakers, the scheme requires two steps namely \textit{enrollment} and \textit{key exchange} to be performed  before the camera can be deployed for \textit{monitoring}. Enrollment, key exchange and monitoring phases of the scheme are depicted in Fig.~\ref{fig:Phases}. The secure node approach uses a trusted authority (TA). Enrollment is performed by the TA in a secure and trusted environment. During this step, the TA extracts a unique fingerprint of the camera hardware using the PUF framework of Section~\ref{subsec:PUF_framework} and binds signing and encryption keys with the hardware. During the key exchange, the caretaker securely transfers the signature verification and decryption keys from the camera device to her monitoring device.  
Afterwards, the camera is deployed for monitoring. To setup our scheme, the TA generates a master key pair $(mpk, msk)\gets K(1^k)$. We assign each camera a unique identity $I$ and a PUF instance. We denote a camera with identity $I$ and on-chip PUF instance $PUF$ as CAM($I, PUF$) and the trusted authority with master key pair as TA($msk,mpk$).
\begin{enumerate}
    \item \textbf{Enrollment.} 
    
    During enrollment, TA($msk$, $mpk$) binds a signing key pair ($sk$, $pk$) and an AES-encryption key ($k_E$) to the camera node CAM($I$,  $PUF$) using the camera fingerprint. The CAM presents TA with its identity $I$ as a request for enrollment. The TA picks two random challenges ($c_1, c_2$) and feeds them to the $PUF$ on the camera. The responses from the $PUF$ ($r_1$, $r_2$) are returned to the TA, who then binds a signing and an encryption key with the CAM as follows:  First, the TA  generates a signing key pair ($sk$, $pk$) using the key generation algorithm of PUF-based Cert IBS. Using the key-binding algorithm of the PUF framework, it binds the private-half of the key pair to the $PUF$ using $r_1$,  i.e., $W_1 \gets Gen(r_1, sk)$. Furthermore, the TA issues a certificate consisting of its signature on the CAM's identity and public half of the key pair, i.e., $cert \gets Sign_{msk}(I, pk)$. Second, the TA generates a unique, random encryption key  $k_E$ and binds it to the $PUF$ on the CAM using $r_2$, i.e., $W_2 \gets Gen(r_2, k_E)$. The tuple ($W_1$,$W_2$, $cert$) is stored in non-volatile memory on camera.
    
    \item \textbf{Key Exchange.}  Key exchange is performed by the caretaker before deploying the camera for monitoring. During this step, the caretaker transfers the signature verification key $pk$ and the decryption key $k_E$ from the camera device to her monitoring device via a local interface such as NFC. Since the transfer is done in a private space using a local connection, it is assumed that $k_E$ is not leaked to a third party. Securing the keys on mobile devices with vulnerable software stack is out of scope of this work. However, well established techniques such as virtualization~\cite{chen2008overshadow} (isolates applications requiring trusted infrastructure) and secure vault can be leveraged for this purpose. 
    
    \item \textbf{Monitoring.} Once the camera is deployed for monitoring, on every power-up, the signing and encryption keys are generated from noisy $PUF$ responses following the key extraction phase of the PUF framework, i.e., $sk\gets Rep(r_1',W_1)$ and $k_E\gets Rep(r_2',W_2)$.  In case of an event of interest, the video footage capturing the event is transferred to the caretaker. Let $N$ be the total number of frames comprising the footage. The value of $N$ can either be a fixed or variable number depending on type of the event. Upon the occurrence of an event, each video frame is encrypted using the AES128 algorithm to ensure confidentiality, i.e., $C_i \gets Enc_{k_E}(frame[i]) |_{\ i\ =\ 1 \ldots N}$. The non-repudiation of the data is ensured by MAC-then-Sign technique. First, each encrypted frame is hashed using the HMAC algorithm to ensure integrity $h_i \gets HMAC(C_i) |_{\ i\ =\ 1 \ldots N}$. This is followed by signing the entire hash-chain of all encrypted frames using the PUF-based Cert-IBS scheme given by $\sigma \gets Sign_{sk}(h_1\bigparallel h_{2} \bigparallel \cdots \bigparallel h_{N} \bigparallel \tau)$ where $\tau$ is the timestamp given by $SHA256(I\parallel event\_count)$. Signing the entire hash-chain together preserves the frame order. The timestamp is included in the signature to ensure freshness of data and thwart replay attacks. The camera then uploads the encrypted-then-MACed-then-signed footage $\{C_1,C_2,\ldots, C_N,\tau, \sigma\}$ to the storage server. An alert message notifies the caretaker about the event and the completion of the upload. The caretaker can then download the footage on-demand and check integrity, authenticity and freshness by verifying the Cert-IBS signature, i.e., $1 \stackrel{?}{=} Ver_{mpk}(cert, (I, pk))$ and 
$1 \stackrel{?}{=} Ver_{pk}(\sigma, (h_1\bigparallel h_{2} \bigparallel \cdots \bigparallel h_{N} \bigparallel \tau))$. Upon successful verification, the caretaker uses the decryption key to decrypt the frames to obtain footage in raw format, $frame[i] \gets Dec_{k_E}(C_i)|_{\ i\ =\ 1 \ldots N}$.
\end{enumerate}

\subsection{Secure Camera Architecture \& Prototype}
\label{SCA}

\begin{figure}
\centering
\subfigure[Zynq7010 SoC and OV5642 image sensor based smart camera platform used for implementation of trusted image sensor and secure camera node approaches]{\includegraphics[width=0.6\columnwidth]{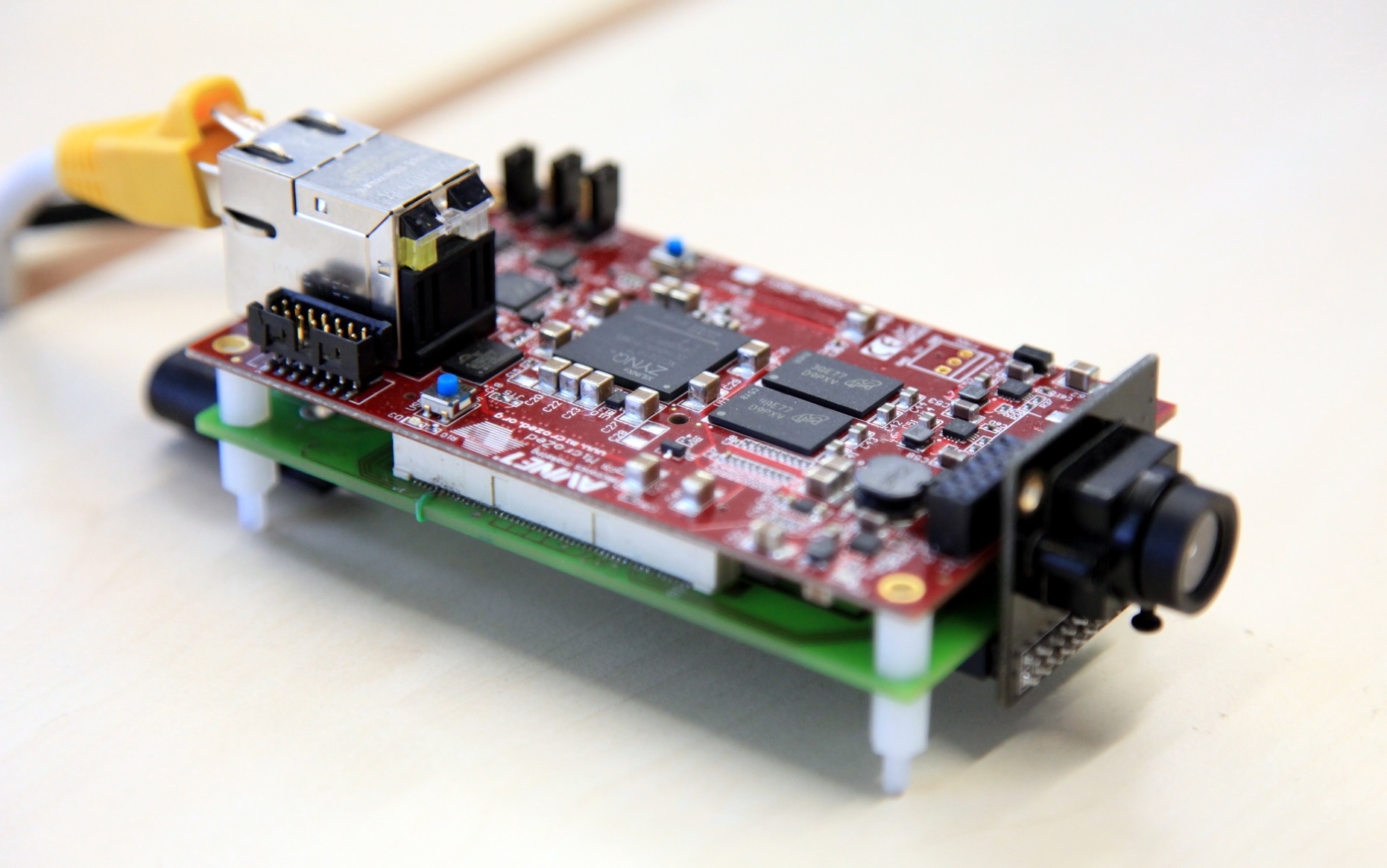}\label{fig:HW}}
\qquad
\subfigure[Block diagram of SoC-based secure camera node that depicts core components of camera hardware (gray) and software tasks (white) performed by these components]{\includegraphics[width=0.6\columnwidth]{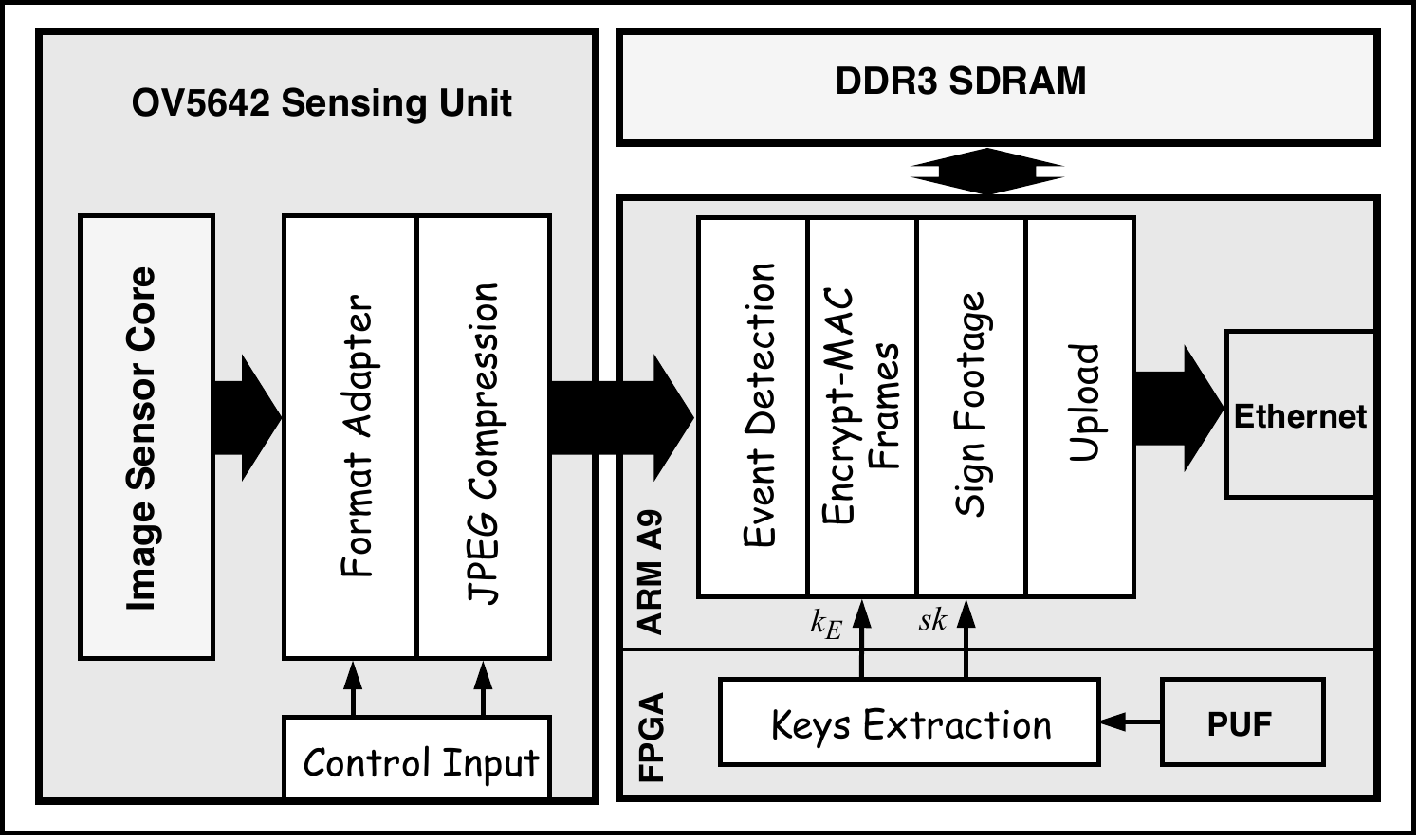} \label{fig:hardware}}
\qquad
\subfigure[Hardware (dark gray) and software (light gray) stack of Zynq SoC-based camera prototype. Hardware comprises of 5MP OV5642 image sensor, Zynq7010 SoC having FPGA and ARM Cortex A9 core. The software stack comprises embedded Linux kernel, system and user libraries and application framework]{\includegraphics[width=0.6\columnwidth]{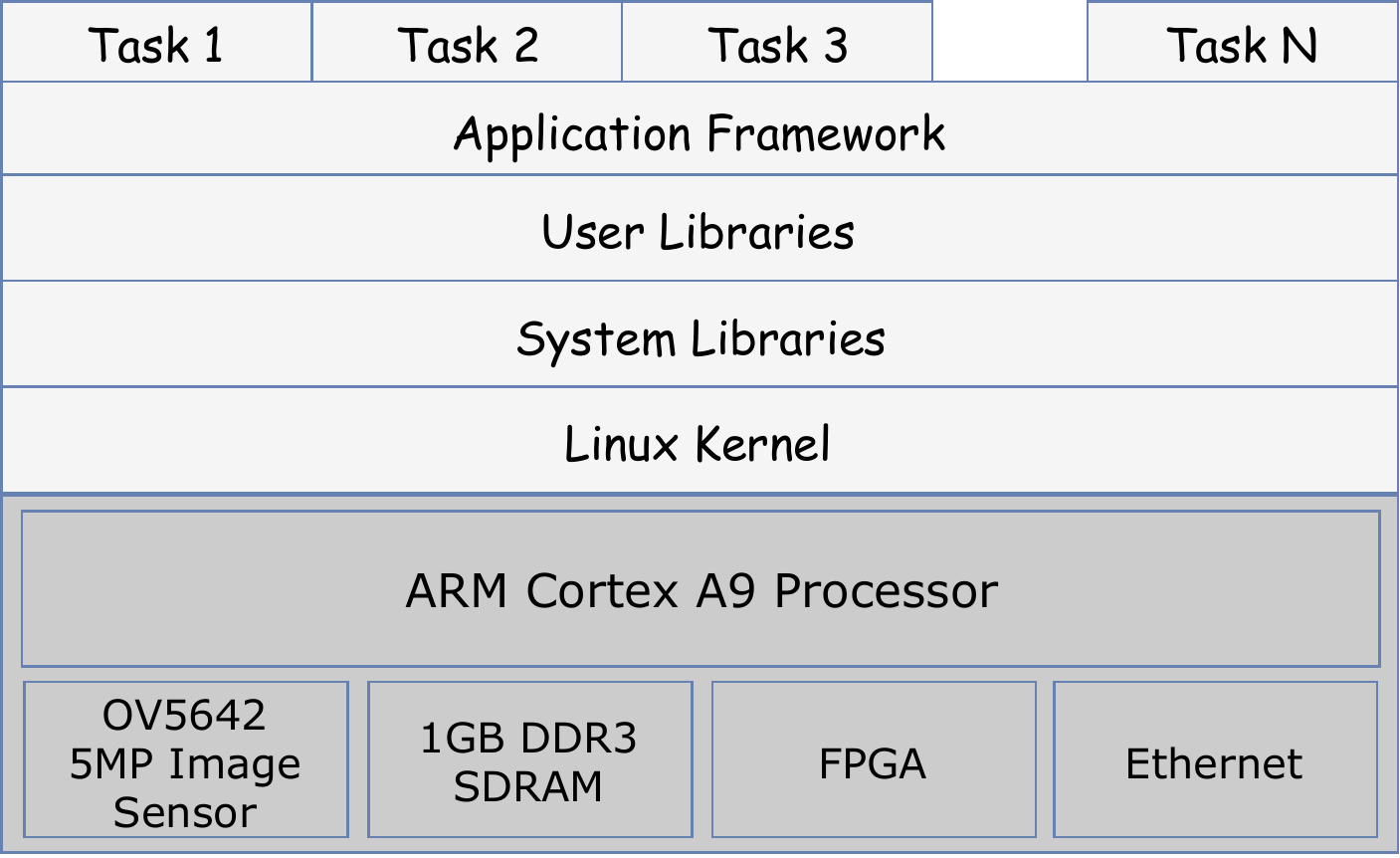} \label{fig:stack}}
\caption{Hardware platform, block diagram, and hardware/software stack of secure camera node}
\label{fig: prototype}
\end{figure}

The key idea underlying the secure camera architecture is to leverage an on-chip PUF to extract the node's fingerprint from the hardware, which serves as basis for on-board data-, node-, and personal privacy protection. Security is rooted in the system hardware making it an intrinsic element of the device and therefore harder to bypass. Video data processing and protection is done inside the SoC and the data leaves the chip with integrity, authenticity, confidentiality, access authorization, and freshness guarantees.

The choice of the processing platform is a critical decision in every vision systems design. CPUs perform operations in sequence whereas FPGAs are massively parallel in nature. Typically, FPGA performs vision processing order of magnitude faster than CPUs. However, an FPGA consumes more power  and has higher programming complexity as compared to a CPU. An architecture featuring both an FPGA and a CPU presents the best of both worlds and often provides a competitive advantage in terms of performance, cost, and power consumption~\cite{CPUvFPGA}. The secure camera node leverages a system-on-chip (SoC). This offers two advantages: First, the SoC provides a monolithic architecture that allows to architect a security solution tightly integrated with the system logic. Second, the SoC comprises an FPGA and processor part which provides the flexibility to compose a security solution addressing all layers of the node stack.  

There exist two categories of operating system (OS), that are used for embedded devices: real-time OS (RTOS) and general-purpose OS. An RTOS provides scheduling guarantees to ensure deterministic behaviour and timely response events and interrupts. However, it is inefficient at handling multiple tasks in parallel and lacks board (hardware) support. Embedded Linux and Android dominate the world of general-purpose OS for embedded systems. Android has been widely successful as mobile OS due to its rich support for multimedia, graphics, user interface, and networking. Drawbacks of Android lie in its large memory footprint and extensive CPU resources consumption. Embedded Linux shines when it comes to operating efficiency in terms of memory footprint, power, and computing performance. The secure camera node uses embedded Linux OS.

\begin{figure*}
\centering
  \includegraphics[width=1.0\linewidth]{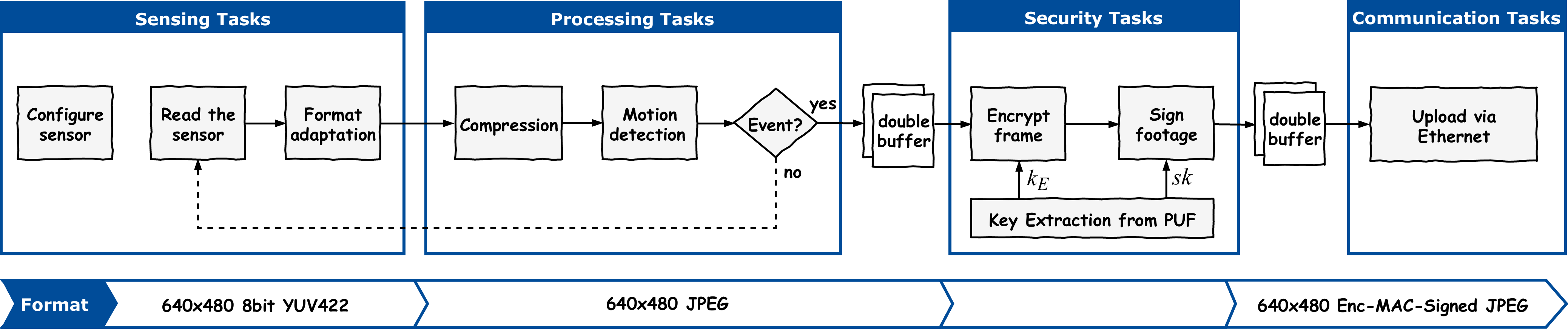}
  \caption{Application framework of the secure camera node comprises sensing, processing, security and communication tasks. The sensing tasks read image data from the sensor and perform format adaptations. The processing tasks include the application logic (e.g., event trigger based on motion detection). In case of an event detection, data is forwarded to security tasks, where frames are encrypted-MACed-signed. Protected frames are forwarded to communication tasks for upload}
  \label{fig:videopath}
\end{figure*}

The prototype shown in Fig.~\ref{fig:HW}, realizes the secure camera node architecture. The block diagram of the node showing the core modules of video data path and their mapping on hardware components is depicted in Fig.~\ref{fig:hardware}. The hardware and software stack of the camera node are depicted in Fig.~\ref{fig:stack}. The camera hardware is comprised of OV5642---a 5 MP CMOS image sensor array, Zynq7010 SoC, 1 GB SDRAM, and a gigabit Ethernet interface. The SoC houses a dual-core ARM Cortex A9 processor clocked at 666 MHz and FPGA fabric. The ARM-A9 processor runs Embedded Linux that hosts system libraries (OpenSSL, GMP, libjpeg etc.) and user libraries (pbc, motiondetection etc.) to be used by the applications. 
On top of the OS, a custom application framework is designed, which is responsible for providing the intended (application specific) functionality to the device. The application framework for the secure camera architecture, given by Fig.~\ref{fig:videopath}, is divided into the four tasks: sensing, processing, security, and communication.

Sensing tasks include reading the visual data from the image sensor and encoding it the desired format. The OV5642 image sensor is configured to provide data in 640$\times$480 (resolution) 8-bit YUV422 (color-space) format. Processing tasks include the application specific logic; the prototype for private space monitoring performs video compression and event detection by behavioral analysis of video data. Video compression is achieved using the JPEG compression engine on the OV5642 sensing unit. The event is triggered if motion is greater than a predefined threshold. Motion of the monitored individual is computed using the three-frame differencing algorithm by Collins et al.~\cite{collins2000system}. The image
difference between frames at time $t$ and $t-1$ and the difference between $t$ and $t-2$, is performed to determine regions of legitimate motion and to erase ghosting.  If an event has been detected, the frames are forwarded to the security tasks. 

Security tasks entail data security and node security. Data security tasks secure the data on-camera as described in Section~\ref{OP}. After camera power-up, encryption and signing keys are extracted from the ring oscillator (RO) PUF implemented using the reprogrammable fabric and are loaded into the cache. Each frame is encrypted using AES128 and MACed using HMAC-SHA256 to ensure data confidentiality and integrity, respectively. MAC checksums of all frames in the footage are concatenated and signed using the PUF-based Cert-IBS scheme with BLS~\cite{boneh2001short} as underlying standard signature scheme; this ensure authenticity and preserves the frames order.  
 
Freshness of data is ensured by including a timestamp $\tau$ before signing the checksums. For timestamp generation, the camera uses an event counter that increments whenever an event is detected. Given that an event is detected by the motion detection algorithm and the event counter is incremented to $event\_count$, then the timestamp is calculated as $\tau =$ SHA256$(I\parallel event\_count)$. A time-stamp value holds true only for a specific event $event\_count$ detected by camera device $I$. Following the event $event\_count$, the footage is timstamped with $\tau$. The value of the timestamp should not repeat among legitimate footages from different events detected by the same camera or among footages from different cameras. This simple check deters replay attacks. It is important to note that encryption and hashing is performed on frames whereas time-stamping and signing is performed on the complete footage. 
 
Node security aims to secure the software and hardware stack of the sensor node; this is achieved  by a secure boot of the SoC and the on-chip PUF. The Zynq7010 SoC provides secure boot functionality as part of its boot procedure that verifies authenticity, integrity and unclonability of the camera's software stack based on digital signatures, message authentication code (MAC), and encryption. 

\begin{figure}
\centering
  \includegraphics[width=0.8\linewidth]{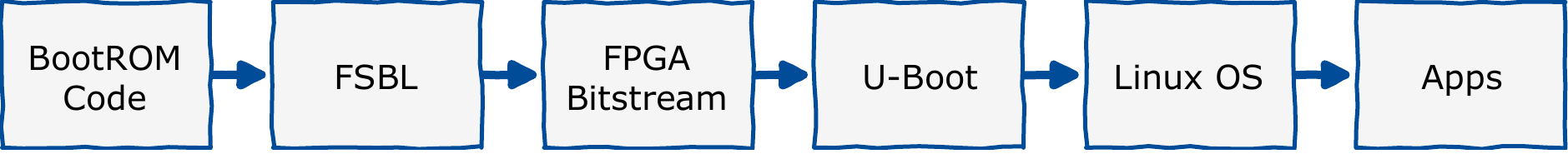}
  \caption{Chain of trust for secure boot of Zynq7010 SoC}
  \label{fig:secureboot}
\end{figure}

The boot mechanism is CPU-driven. Other hardware components used in the boot process are the non-volatile memory (NVM), BootROM, on-chip memory (OCM), AES/HMAC module, JTAG, and DDR RAM. The software programs involved in the boot are the BootROM code, the first-stage bootloader (FSBL), U-Boot, the Linux kernel, and user applications. Boot-chain of the camera device is depicted in Fig.~\ref{fig:secureboot}. The foundation of secure boot is established by placing the BootROM code in a mask ROM, a one-time programmable memory, which implies that the ROM contents cannot be modified. While creating the image for secure boot, each successive component of the boot-chain is signed (RSA), hashed (HMAC-SHA256), and encrypted (AES256). The boot up starts with the BootROM code loading the FSBL, and continues serially with the FSBL loading the FPGA bitstream and the software. During every secure boot up, the chain of trust is established by the successive verification of signature (authentication), MAC checksum (integrity) and decryption (confidentiality) of all software, i.e., FSBL, bitstream, u-boot, OS, and user applications. 

This procedure prevents an adversary from tampering with software or the FPGA bitstream file. Zynq SoC contains hard IP cores for AES decryption and HMAC computation. As a result, the difference between the boot time of secure  and regular boots is negligible~\cite{sanders2013secure}. The hardware stack is protected by the on-chip PUF. Incorporating a PUF into a chip makes the chip tamper evident~\cite{roel2012physically}. Since PUF behavior corresponds to the underlying silicon fabric, any tampering with the fabric modifies the PUF behavior, thereby modifying the camera fingerprint. This leads to generation of incorrect signing and encryption keys thereby incorrect signature and cipher text, which is detected by the verifier. 

Data with confidentiality, integrity, authenticity and freshness guarantees is then forwarded to communication module for uploading. The prototype merely demonstrates a proof of the concept and can be extended with wireless communication capabilities such as WiFi.  

\section{Implementation \& Evaluation}
\label{implementation}

The section evaluates the \textit{trusted sensing and privacy-aware reporting} and \textit{secure node} approaches for IoT-based smart services. The section is divided into five parts: Since both approaches utilize the PUF framework to generate and store sensor bound keys, we present the implementation results of the PUF framework in Section~\ref{implement: PUF}. The \textit{privacy-aware trusted sensing} scheme is evaluated in two parts: First, Section~\ref{OverallSensor} presents the trusted image sensor prototype and evaluates the overhead on the sensor with respect to storage, latency and hardware. Second, Section~\ref{eff} evaluates the communication overhead on the mobile device for \textit{privacy-aware reporting} of the sensed data from trusted sensors. Section~\ref{SecNodeImpl} evaluates the storage, latency, hardware, and communication overhead on the secure camera node (the prototype presented in Section~\ref{SCA}) incurred due to \textit{secure node} approach. Section~\ref{SecProp} discusses security properties and limitations of both schemes.  

The experimentation setup used for the evaluation of the privacy-aware trusted sensing and the secure node approaches is as follows: In order to verify the feasibility of a PUF-based approach for sensors, we implemented the PUF framework on three sensing platforms of different complexities, namely (i) Atmel ATMEGA328P, a lightweight 8-bit MCU running at 8 MHz,  (ii) ARM Cortex M4, a 32-bit MCU running at 168 MHz, and (iii) Xilinx Zynq7010 SoC with FPGA and a 32-bit dual core ARM Cortex A9 processor core running at 666 MHz. For objective comparison of the privacy-aware trusted sensing and the secure node approaches, we prototyped a trusted image sensor and a secure camera node using the same platform as shown in Fig.~\ref{fig:HW}. The platform comprises a 5MP OV5642 image sensor module and MicroZed board, which houses Zynq7010 SoC clocked at 666MHz, 1 GB external RAM for frame buffering and a gigabit Ethernet interface to upload video footage. A custom board (mounted below the Micozed board) was designed to interface the image sensor with the MicroZed board and regulate power to both the modules.

\subsection{PUF Framework}
\label{implement: PUF}

Various PUF sources are inherent to a typical sensor including SRAM PUF, RO PUF, and sensor-specific PUFs~\cite{cao2015cmos, rosenfeld2010sensor, rajendran2016securing}. Since we target a broad range of sensors, we seek to identify PUF sources that are commonly available on most sensors such as SRAM and RO PUFs.

A PUF is characterized by three quality parameters, i.e., randomness, reliability,  and uniqueness. Hamming weight ($HW$), the indicator of PUF randomness, measures the deviation of a PUF output from uniform distribution. For an $n$ bit response $r$ obtained from a chip $U$, $HW$ is given by:

\begin{equation}\label{eq:HW}
HW(r)=\frac{1}{n} \sum\limits_{i=1}^{n} b_{i}\cdot 100\ \%
\end{equation}
\noindent
where $b_i$ is the $i^{th}$ binary bit in an $n$ bit PUF response. For a uniformly distributed PUF response, $HW$ should be $50\ \%$.

The change in PUF response over varying environmental and operating conditions depicts the (lack of) reliability. The change is referred to as PUF error-rate or noise and is measured in terms of the inter-Hamming distance.  An $n$ bit reference response ($r_{ref}$) is extracted from the chip $U$ at the room temperature and standard operating conditions.  Multiple responses from the same PUF are obtained under different environmental and operating conditions (e.g. varying temperature or supply voltage) and are denoted by $r_i$. A number of samples of $r_i$ are taken for each combination of the environmental and operating conditions. The PUF error-rate is measured as the average intra-Hamming distance ($HD^{intra}$) over $N$ samples obtained under different environmental and operating conditions. For the chip $U$, it is defined as:

\begin{equation}\label{eq:HDintra}
HD^{intra}(r_{ref}, r_{1}, r_{2}, \cdots , r_{N-1} )= 
\frac{1}{N}\sum\limits_{i=1}^{N}  \frac{HD^{intra}(r_{ref}, r_{i})}{n} \cdot 100\ \%
\end{equation}

The uniqueness property of a PUF measures how unique are the signatures generated from different
chips using the same PUF circuit. The average inter-Hamming distance ($HD^{inter}$) of the PUF responses is a commonly used measure of uniqueness. The average $HD^{inter}$ for a group of $M$ chips is defined as the average of all possible pair-wise $HD^{inter}$ among $M$ chips:

\begin{equation}\label{eq:HDinter}
HD^{inter}(U_1, U_2, \cdots,U_M)= \frac{2}{M(M-1)}\sum\limits_{U_1=1}^{M-1} \sum\limits_{U_2=U_1+1}^{M} \frac{HD^{inter}(r_{U_1}, r_{U_2})}{n} \cdot 100\ \%
\end{equation}

For a truly random PUF output, $HD^{inter}$ should be close to $50\ \%$.

We implemented PUFs on three platforms: (i) Atmel ATMEGA328P, a lightweight $8$ bit MCU,  (ii) ARM Cortex M4, a $32$ bit MCU, and (iii) Xilinx Zynq7010 SoC with re-programmable logic and a dual core ARM Cortex A9. These platforms are ideally suited as sensor controllers (see Fig.~\ref{fig: SolScn1}) for a broad range of sensors.

SRAM PUF taps the randomness from the start-up values of the SRAM cells. Once these start-up values are read out, the SRAM can be used as regular memory. As a result, SRAM PUF implementation does not incur any hardware overhead and is therefore preferred for implementation over the RO PUF. However, if SRAM is either not available on board or gets initialized with fixed values during boot up, the RO PUF is implemented. The power-up state of the SRAM cells on the ATMEGA328P and the ARM Cortex M4 show PUF behavior where as the SRAM on the Zynq7010 SoC gets initialized with fixed values during boot up. Therefore, we implemented the SRAM PUF on the ATMEGA328P and the ARM Cortex M4 MCUs and the RO PUF on the Zynq7010 SoC.

\begin{figure*}
\centering
\subfigure[]{\includegraphics[width=.7\linewidth]{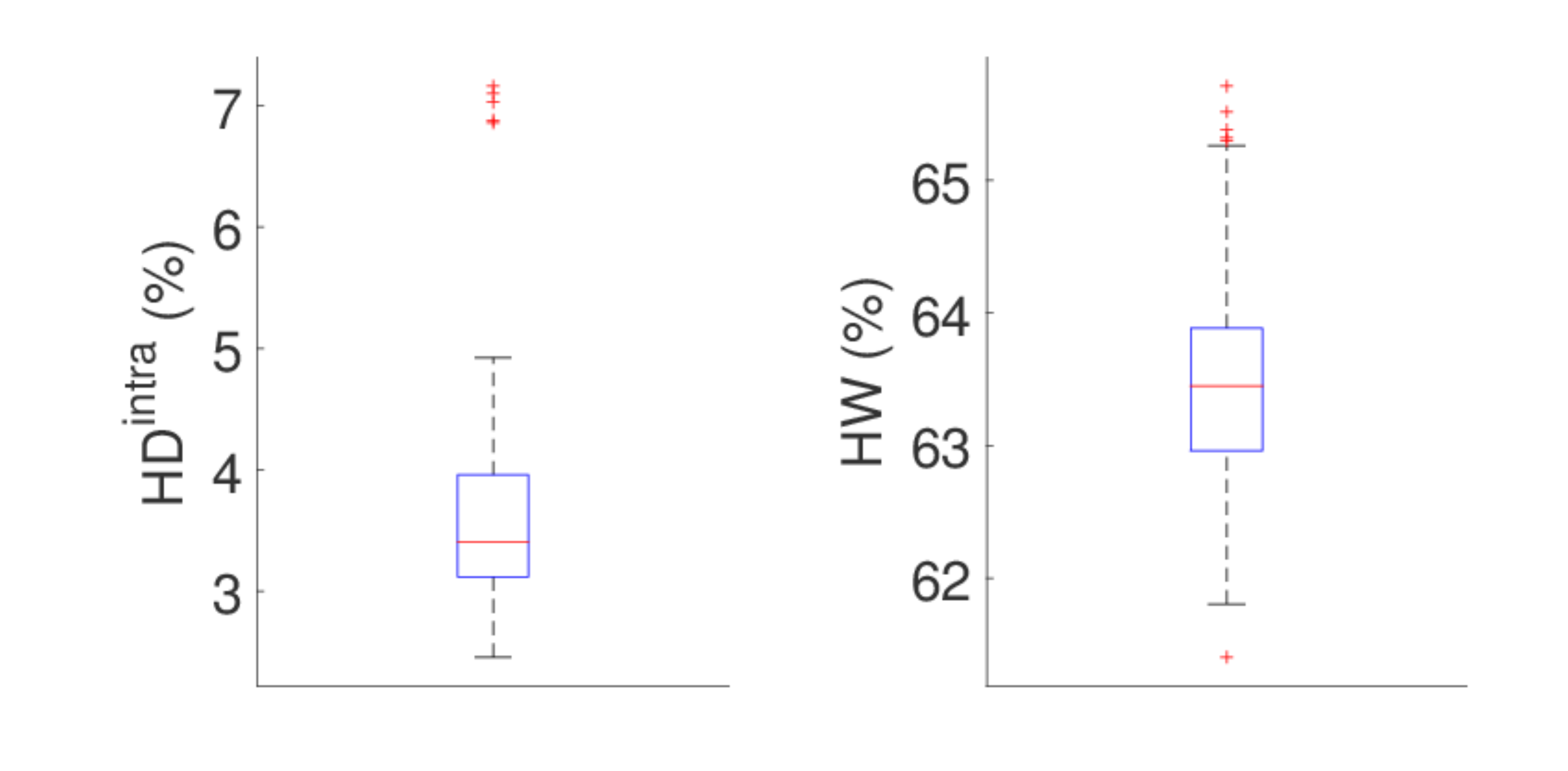}\label{fig:SRAM_MCU8}}
\quad    
\subfigure[]{\includegraphics[width=.7\linewidth]{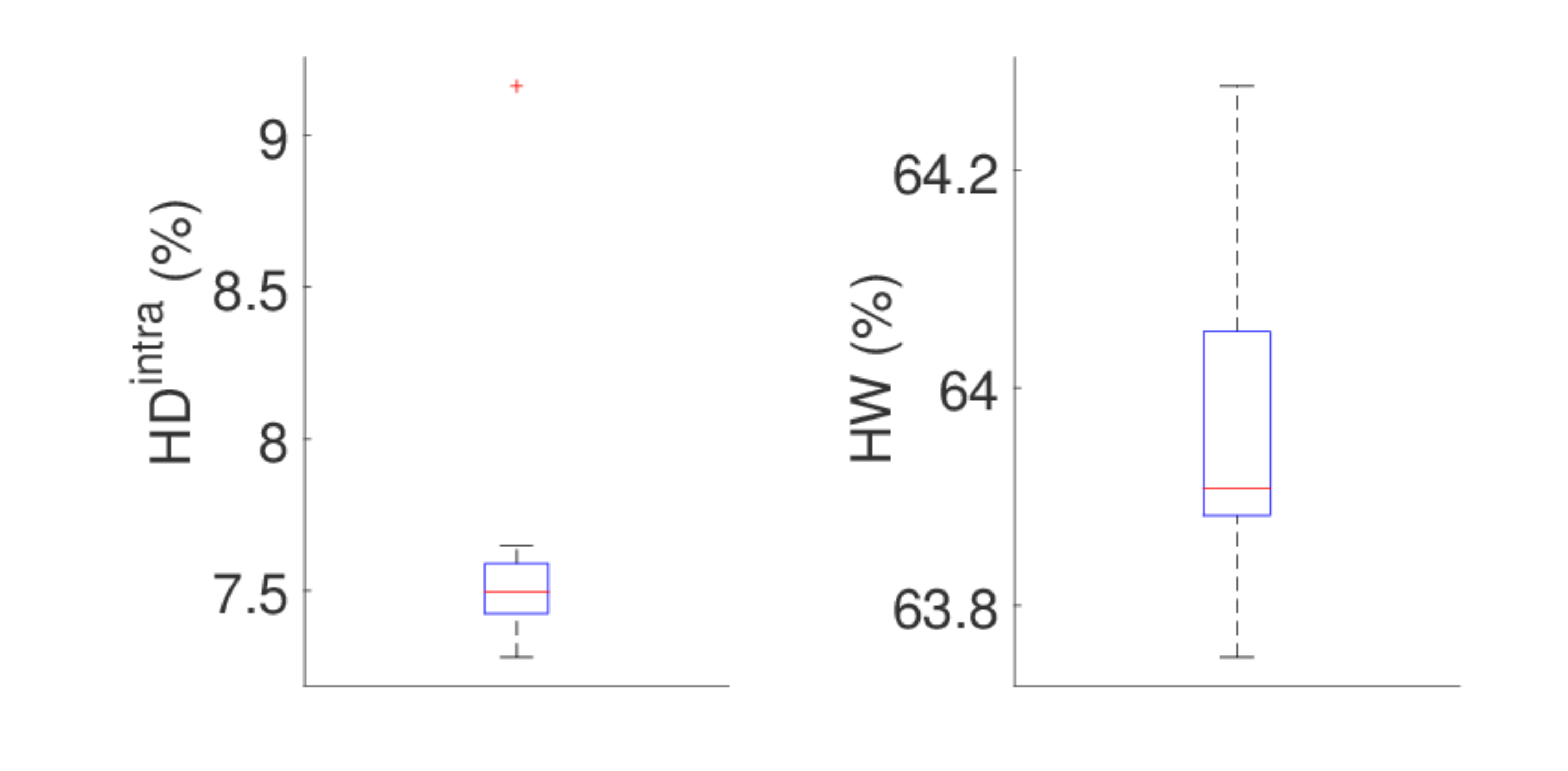}\label{fig:SRAM_ARMCortexM4}}
\quad    
\subfigure[]{\includegraphics[width=.7\linewidth]{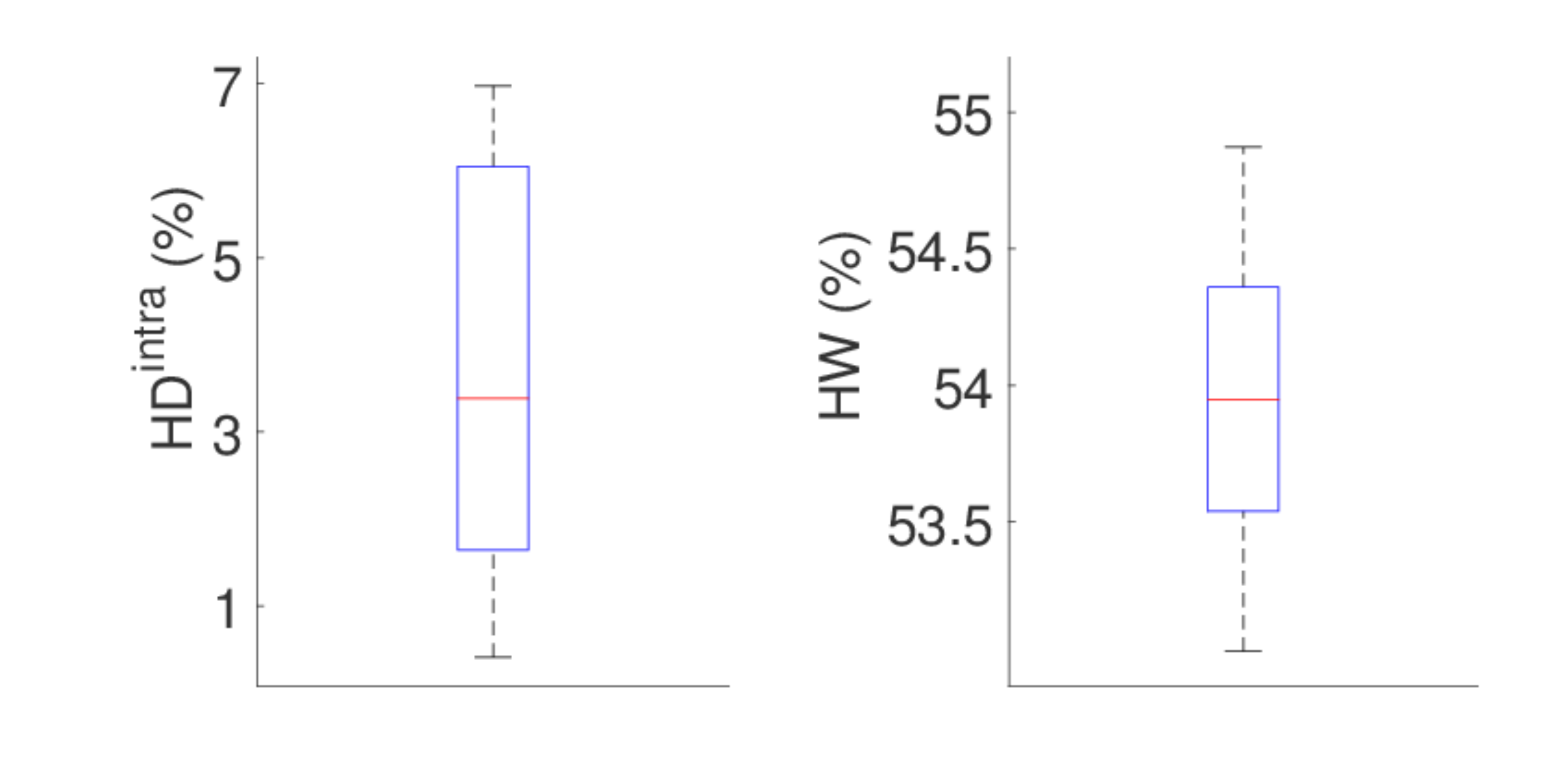}\label{fig:RO_Zynq7010}}
\caption{PUF characterization of (a) $1 kB$ SRAM on Atmel ATMEGA328P 8-bit MCU, (b) $15 kB$ SRAM on ARM Cortex M4 32-bit MCU, and (c) RO PUF comprised of $1040$ 3-stage ring oscillators implemented on re-programmable fabric of Xilinx's Zynq7010 SoC. The PUF quality parameters for (a) and (b) are calculated over 100 PUF responses at room temperature. The mean and maximum values of the intra-Hamming distance $(\mathrm{HD^{intra}})$ representing the PUFs' error-rate for (a) are $3.4 \%$ and $7.2 \%$, and for (b) $7.66\%$ and $9.16\%$, respectively. The randomness of PUF responses, measured as the mean Hamming weight $(\mathrm{HW})$, for (a) and (b) are $63.5 \%$ and $63.96 \%$, respectively. The quality parameters for (c) are calculated from a total of 800 responses taken over temperature range $0-60^{\circ}\mathrm{C}$ (i.e., 100 responses at $10^{\circ}$ intervals plus a 100 at $25^{\circ}\mathrm{C}$) where a response $\approx1560$ bits. The mean and maximum values of $\mathrm{HD^{intra}}$ are $3.6 \%$ and $6.97 \%$, respectively. The average $\mathrm{HW}$ is $53.95 \%$} 
\label{fig:PUFidentification}
\end{figure*}

For the SRAM PUF implementation, $1$ kilobytes SRAM on the Atmel ATMEGA328P and $15$ kilobytes SRAM on the ARM Cortex M4 were read out and characterized for PUF behavior. The PUF quality parameters were computed from $100$ PUF responses (i.e., start-up values of the SRAM cells) obtained at room temperature.  Figs.~\ref{fig:SRAM_MCU8} and ~\ref{fig:SRAM_ARMCortexM4} depict the error-rate and the uniform distribution of PUF-responses measured as $HD^{intra}$ and $HW$, respectively. The average and maximum values of $HD^{intra}$ were measured as $3.4\ \%$ and $7.2\ \%$ for the ATMEGA328P, and $7.66\ \%$ and $9.16\ \%$ for the Cortex M4. The randomness of the PUF responses for the ATMEGA328P and the Cortex M4 were measured as $63.5\ \%$ and $63.96\ \%$, respectively. 

Furthermore, we implemented the RO PUF, comprised of $1040$ 3-stage ring oscillators and two $16$ bit counters, on the FPGA part of the Zynq7010. $1040$ ROs can be arranged  into $1039$ RO independent pairs. To obtain a PUF response, a RO pair is selected. The frequencies generated by the selected ROs, increment the respective counters. The marginal difference between the two frequencies causes one counter to overflow before the other. At the overflow of one counter, the $16$ bit value of the other counter is read out. Three bits at positions 8, 9, and 10 in the counter value are read out as the output of the RO pair comparison~\cite{kodytek2015newPUF}. Therefore, $1039$ RO pairs generate a $3117$ bit response. We evaluated the RO PUF for $HD^{intra}$, $HW$ and $HD^{inter}$.  The quality parameters for the Zynq7010 were computed from a total of $800$ responses measured over a temperature range of $0-60^{\circ}\mathrm{C}$ (i.e., $100$ responses at $10^{\circ}$ intervals, and an additional $100$ responses at $25^{\circ}\mathrm{C}$). The average and maximum $HD^{intra}$ were computed as $3.6\ \%$ and $6.97\ \%$, respectively. The average $HW$ was $53.95\ \%$. Furthermore, the same PUF was implemented on $10$ Zynq7010 boards, and $HD^{inter}$ was $51.1\ \%$. 

Instead of designing three separate helper data algorithms (HDAs) for correcting $7.2\ \%, 9.16\ \%,$ and $6.97\ \%$ error-rates of the three PUF, we designed a common HDA that can correct an error-rate of up to $10\ \%$. Guajardo et al.~\cite{guajardo2007physical} investigated different error correcting codes that are suitable for HDA. The error-correcting code determines the number of required PUF response bits and hence the size of PUF. We evaluated two cases: (i) a simple code using BCH (492,57,171) and (ii) a concatenated code comprising Reed Muller (16,5,8) and Repetition (5,1,5) codes. The failure rate for PUF-based key reconstruction using both these cases is $\leq 10^{-6}$. 

Once an error correcting code $(n,k,d)$ is selected, the  number of PUF response bits required to generate an $l$ bit key is given by $\frac{n}{k}\cdot l$. According to the PUF framework, the helper data $W$ has the same length as the PUF response. Since, an RO pair generates $3$ response bits, $(\frac{n}{k}\cdot l)$ bits are generated by $(\frac{n}{3k}\cdot l)$ RO pairs or $(\frac{n}{3k}\cdot l+1)$ ROs. In our design, each RO is implemented as 3-stage (3 NOT gates), the RO PUF's hardware size is given by $(\frac{n}{k}\cdot l+3)$ logic-gates. For a concatenated code $(n_1,k_1,d_1)||(n_2,k_2,d_2)$, the hardware and helper data sizes can be computed by using the same procedure, with $n=n_1$ and $k=k_2$. 

In the \textit{trusted sensing and privacy-aware reporting} approach, a sensor uses $160$ bit $sk$ to sign the sensor readings (sensed data attestation) using PUF-based Cert-IBS with BLS as underlying signature scheme  where as the \textit{secure node} approach uses a $128$ bit $k_E$ to encrypt the frames using AES128 algorithm and a $160$ bit $sk$ to sign the footage using the PUF-based Cert-IBS with BLS as underlying signature scheme. 
The PUF framework was implemented on all three platforms. In Table~\ref{tab:eval} we present the overhead incurred by the PUF framework to generate $128$ and $160$ bit keys, the latency, hardware and memory components.   

\begin{table}[h]

\centering
\resizebox{0.85\columnwidth}{!}{
\begin{tabular}{|c|c|c|c|c|c|}
\hline
\rule{-2pt}{4ex}
$\mathrm{\mathbf{PUF\ Source}}$ & $\mathrm{\mathbf{Key\ Length}}$  & $\mathrm{\mathbf{Error\ Correcting}}$               & $\mathrm{\mathbf{Hardware}}$         & $\mathrm{\mathbf{Latency}}$              &  $\mathrm{\mathbf{Storage}}$  \tabularnewline
                                &                                 & $\mathrm{\mathbf{Code}}$                            &  ($\approx$ Logic Gates)         & (Key Extraction)                     &  (Helper Data $W$)            \tabularnewline
\hline
\rule{-2pt}{3ex}
\multirow{4}{*}{$\mathrm{\mathbf{RO}}$}       & \multirow{2}{*}{128-bit}     & $\mathrm{{BCH}}$                                    &  $1108$                            &                       &  $\mathbf{1105}$ bits          \tabularnewline
\cline{3-4} \cline{6-6}
\rule{-2pt}{3ex}                                                   
                                       &                                             & $\mathrm{{RM\ ||\ Rep}}$                            &  $2051$                            &     $\leq 100\ \mathrm{ms}$                                     &    $2048$ bits                 \tabularnewline
%\hline
 \cline{2-4} \cline{6-6}
\rule{-2pt}{3ex}
                                       & \multirow{2}{*}{160-bit}             & $\mathrm{{BCH}}$                                    &  $1384$                            &                   &  $\mathbf{1381}$ bits          \tabularnewline
\cline{3-4} \cline{6-6}
\rule{-2pt}{3ex}                                                   
                                       &                                             & $\mathrm{{RM\ ||\ Rep}}$                            &  $2563$                            &                                          &    $2560$ bits                 \tabularnewline
\hline
%\hline
%\rule{-2pt}{4ex}
%$\mathrm{\mathbf{PUF\ Source}}$ & $\mathrm{\mathbf{Key}}$  & $\mathrm{\mathbf{Error\ Correcting}}$               & $\mathrm{\mathbf{Hardware}}$         & $\mathrm{\mathbf{Latency}}$              &  $\mathrm{\mathbf{Storage}}$  \\
%                                &                                 & $\mathrm{\mathbf{Code}}$                            &  ($\approx$ Logic Gates)         & (Key Extraction)                     &  (Helper Data $W$)            \\
%\hline
\rule{-2pt}{3ex}
\multirow{4}{*}{$\mathrm{\mathbf{SRAM}}$}       & \multirow{2}{*}{128-bit}     & $\mathrm{{BCH}}$                                    &  NA                            &                       &  $\mathbf{1105}$ bits          \\
\cline{3-4} \cline{6-6}
\rule{-2pt}{3ex}                                                   
                                       &                                             & $\mathrm{{RM\ ||\ Rep}}$                            &  NA                            &     $\approx 30\ \mathrm{ms}$                                     &    $2048$ bits                 \\
%\hline
 \cline{2-4} \cline{6-6}
\rule{-2pt}{3ex}
                                       & \multirow{2}{*}{160-bit}             & $\mathrm{{BCH}}$                                    &  NA                            &                   &  $\mathbf{1381}$ bits          \\
\cline{3-4} \cline{6-6}
\rule{-2pt}{3ex}                                                   
                                       &                                             & $\mathrm{{RM\ ||\ Rep}}$                            &  NA                            &                                          &    $2560$ bits                 \\
\hline
\end{tabular}}
\vspace{3mm}
\caption{Implementation results of the PUF framework for $128$ bit and $160$ bit keys generation and storage}
\label{tab:eval}
\end{table}

\subsection{Trusted Image Sensor}
\label{OverallSensor}

The trusted sensor  prototype (Fig.~\ref{fig: SolScn1}) comprises an $OV5642$ image sensor array (sensing unit) and Zynq7010 SoC (sensor controller) running at $666$ MHz. The sensing unit was configured for $640\times480$ resolution and $YUV422$ color-space. The evaluation matrix for the trusted image sensor comprises three components of the overhead incurred due to the PUF-based Cert-IBS, i.e., storage, latency, and hardware.

During the enrollment phase of the privacy-aware trusted sensing scheme, TA binds a signing key $sk$ to the trusted sensor using the PUF framework and provides a certificate on the public key $pk$. Using the asymmetric version of the BLS signature scheme as the $SS$ in the PUF-based Cert-IBS, we obtain the sizes of $sk=160$ bits and $cert=480$ bits ($pk=160$ bits, $Sign_{msk}=320$ bits). The RO PUF with BCH error correcting code based framework (summarized in Table~\ref{tab:eval}) was implemented on the FPGA part of the Zynq7010 SoC. The PUF-based secure key generation and storage framework for a $160$ bit $sk$ incurs $1381$ bits of memory for the storage of helper data ($W$) and $1384$ logic-gates of hardware overhead. 

During the trusted sensing and privacy-aware reporting phase, the trusted sensor performs the sensed data attestation. During this step, a fresh image frame is read from the $OV5642$ image sensor, a MAC checksum is computed over the frame using HMAC-SHA256 algorithm. The checksum is then signed using the PUF-based Cert-IBS with asymmetric version of BLS as the underlying signature scheme. The sensor then outputs the tuple $(M,I,\sigma, cert)$. 

The storage, hardware and latency overhead incurred by the privacy-aware trusted sensing approach on the sensor is summarized as follows:

\begin{itemize}

\item \textbf{Storage.} The trusted sensor stores helper data $W$ and certificate $cert$. For an $l$ bit key, the helper data $W$ size is given by $\frac{n}{k}\cdot l$, where $n$ and $k$ are parameters of the chosen error correcting code. For a $160$ bit key generation framework using  $BCH\ (492,57,171)$, $l=160$ bits, $n=492$ and $k=57$, which gives the size of $W=1381$ bits. The $cert$ is given by $480$ bits. This amounts to a total of $233$ bytes of storage overhead.

\item \textbf{Latency.} Key extraction using the PUF framework is performed only at the start-up and therefore run-time latency overhead is only incurred by the sensed data attestation phase of PUF-based Cert-IBS scheme. During this phase, the sensor controller obtains a new frame from the image sensor, computes a MAC checksum over the frame and signs the checksum using PUF-based Cert-IBS with BLS as underlying signature scheme. \textit{Pairings based cryptography} library~\cite{pbc} was leveraged for implementation of sensed data attestation. The sensor controller, Zynq7010 SoC, requires $2.5\ \mathrm{ms}$ to MAC a frame and  $6.27\ \mathrm{ms}$ to sign the MAC at $640\times480$ resolution and YUV422 color-space, which enables the prototype trusted image sensor to secure $114$ frames per second. 

\begin{comment}
\begin{table}[h]
\centering
	\scalebox{0.9}{%
        \begin{tabular}{l|c|c}
			& $\mathrm{\mathbf{Signing\ Time}}\ (ms)$ & $\mathrm{\mathbf{Verification\ Time}}\ (ms)$\\
			\hline
			%\multirow{5}{*}{\begin{sideways}\scalebox{.9}[1]{\textbf{Code-Offset}} \end{sideways}} & 
            \rule{-2pt}{3ex}
            $\mathrm{\mathbf{Symmetric}}$ & $62$ & $53$\\
            \hline
            \rule{-2pt}{3ex}
            $\mathrm{\mathbf{Asymmetric}}$&  $\mathbf{6.27}$  & $638.19$\\
     		%\hline
	    	\end{tabular}}
		    %\vspace{2mm}
		    \caption{Signing and verification times for BLS signatures performed on ARM Cortex A9 running GNU/Linux 4.4.0 kernel clocked at 333MHz for symmetric ($G\approx 520\ bits$ and $G_T\approx1024\ bits$) and asymmetric settings ($G_1\approx160\ bits, G2\approx320\ bits$ and $G_T\approx1920\ bits$). }
		    \label{table:SDA}
            \end{table}
\end{comment}            

\item \textbf{Hardware.}  The hardware overhead incurs only in case of an RO PUF implementation and results in $1384$ logic-gates of hardware overhead to generate a $160$ bit signing key (cp.~Table~\ref{tab:eval}).

\end{itemize}

\subsection{Privacy-Aware Reporting}
\label{eff}

The evaluation of privacy-aware reporting of the trusted sensors readings aims to compute the communication overhead incurred by the proposed scheme on the user devices. For ease of description, the \textit{trusted sensing and privacy-aware reporting} scheme was explained using the symmetric pairings settings (Table~\ref{FullScheme}). However, symmetric pairing can only be realized using supersingular elliptic curves. Supersingular elliptic curves $E$ are defined over the finite field $\mathbb{F}_q$, where $G$ and $G_T$ are the groups of elliptic-curve points on $E(\mathbb{F}_q)$ and $E(\mathbb{F}_{q^d})$, respectively and $d$ is called the embedding degree of $E$. After recent successful attacks on supersingular curves of small characteristic~\cite{cryptoeprint:2013:400}, the available supersingular curves, given by $y^2=x^3+x$ over the field $\mathbb{F}_q$ for some prime $q=3\ (\mathrm{mod}\ 4)$, have a small embedding degree of $2$. This implies that the base field $\mathbb{F}_q$ must be large enough to obtain sufficient discrete-log security in $\mathbb{F}_q^2$. For instance, to obtain $1024$ bit discrete-log security in $\mathbb{F}_{q^2}$ ($\lceil log_2{q^2}\rceil\geq 1024$), $q$ must be at least 512 bits ($\lceil log_2{q}\rceil\geq 512$). Since $G \subseteq E(\mathbb{F}_q)$, this gives us the size of a group element in $G$ to be $512$ bits. Therefore, the communication overhead on a mobile device incurred by our scheme in the symmetric settings is given by $(6Q+12)512$ bits. 
 
However, the same level of security can be achieved by more efficient curves using the asymmetric pairings setting. Therefore, we have implemented the proposed scheme using asymmetric pairings setting. First, we briefly outline our scheme in the asymmetric settings to compute the overhead in terms of group elements and then we select the curves to compute the communication overhead on the mobile device. 
 
 In the asymmetric setting, the PUF-based Cert-IBS uses asymmetric version of BLS signature scheme. Here, the size of a signature, i.e., $Sign_{sk}(M)$ and $Sign_{msk}(I,pk)$ is given by one element in $G_1$ whereas the public half of the signing key, i.e., $pk$ is given by one element in $G_2$. Proof generation uses the same  framework, $P_\mathrm{NIWI}$, but follows the asymmetric construction based on the $\mathrm{SXDH}$ assumption (cp.~Section 9 of~\cite{groth2012efficient}). Here, commitment to the elements of witness in $G_1$ and $G_2$ costs ${G_1}^2$ and ${G_2}^2$, respectively. The proof consists of two parts: $\pi \in {G_1}^4$, and $\theta \in {G_2}^4$. Given $Q$ sensors' readings with all signatures aggregated into $\bar{\sigma}$, the prover $P_{NI}$ of $P_{NIWI}$ outputs $(\{\mathbf{d}(I_i),\ \mathbf{c}(pk_i)\}_{i=1}^Q,\ \mathbf{c}(\bar{\sigma}), \pi, \theta))$ which amounts to an overhead of $(6+2Q)G_1 +(4+2Q)G_2$. 

Barreto-Naehrig (BN) curves~\cite{barreto2005pairing} are ideal for an asymmetric setting as they offer much higher security with more efficient curves. An efficient BN curve \cite{pbc} allows us to represent elements of $G_1$ with $160$ bits, $G_2$ with $320$ bits and $G_T$ with $1920$ bits. This offers $1920$ bit of discrete-log security in $G_T$. In the asymmetric setting, the communication overhead is given as $(6+2Q)160+(4+2Q)320$ bits. 

For a concrete comparison, we evaluated the communication overhead for two real-word participatory sensing (PS) applications:  $\mathrm{Google\ Street\ View}$~\cite{GoogleSV} and $\mathrm{Wikicity}$~\cite{WikiRome}. The former models a typical PS application that requires its participants to submit multiple sensor readings. Moreover, the payload includes multimedia data.  The later models the worst-case scenario with respect to overhead, because it requires its participants to submit only a single sensor reading (i.e., scalar value). 
%that is, minimum overhead incurred by the the scheme on the trusted sensor.

$\mathrm{APP\ 1:\ Google\ Street\ View}$~\cite{GoogleSV} which requires the participants to capture and upload geotagged images to the Google Maps using their smartphone (user device) cameras and GPS receivers (sensors). Here $Q=2$, where $M_1$ is image from the camera and $M_2$ location reading from the GPS receiver. Given a sensor configuration of $640\times 480$ resolution, RGB color-space, $8$ bits per color-plane, image size $M_1=900$ kilobytes is given. Typically, the GPS receiver provides location data in NMEA format. The maximum length of an NMEA sentence ($M_2$) is $82$ bytes. Therefore, the total size of the payload (i.e., $M_1+M_2$) is given by $900$ kilobytes (approximately). The communication overhead incurred on a smartphone for reporting ($M_1+M_2$)to $\mathrm{Google\ Street\ View}$ server amounts to $520$ bytes that is $0.00056\ \%$ of the payload.

$\mathrm{APP\ 2:\  Wikicity}$~\cite{WikiRome} is an urban planning application that periodically captures the location of the citizens leveraging smartphones and vehicles-embedded location sensors, to monitor their reactions to various events happening in the city. Here, $Q=1$ and $M_1$ is $82$ bytes of location data. For $\mathrm{Wikicity}$, the communication overhead amounts to $400$ bytes. $\mathrm{APP\ 1}$ models a typical multimedia report where as $\mathrm{APP\ 2}$ determines the minimum communication overhead incurred by the privacy-aware trusted sensing scheme on a user device.

\subsection{Secure Camera Node}
\label{SecNodeImpl}

This section presents the implementation results of the prototype secure camera node as discussed in Section~\ref{SCA}.  The node uses a $128$ bit key, $k_E$, to encrypt each frame using AES128 and a $160$ bit key, $sk$ to sign the footage using BLS signature scheme. A RO PUF and BCH error correcting code based framework (Table~\ref{tab:eval}) was implemented for the generation of $sk$ and $k_E$.   The total overhead in terms of latency, storage, hardware and communications incurred due to sensing, processing, security and communication tasks of the \textit{secure node} approach on the camera node is given below.  

\begin{itemize}

\item \textbf{Latency.} Since key extraction and secure boot is performed only once at power-up, they do not incur latency during runtime.  Time-stamping and signing are performed once per footage. At a resolution of $640\times480$, the runtime latency due to the application framework was measured as $27$ ms (i.e., $37$ frames per second). However, at the given resolution, the image sensor outputs $30$ frames per second, which limits the throughput of the prototype to $30$ frames per second. The running times measured for the individual tasks of the camera application framework are given in Table~\ref{table:runtimes}. 

\begin{table}%[h]

	\centering
	\begin{tabular}{lc}
	\toprule
    	\textbf{Module} & \textbf{Runtime}\\
	\toprule
	    \rule{-2pt}{3ex}
        Keys Extraction (PUF Framework)  & $<$100 ms \\
		Event Detection & 21.1 ms \\
		Frame Encyption (AES128)$^\dagger$ & 3.4 ms \\ 
		HMAC-SHA256$^\dagger$ & 2.5 ms \\ 
		Footage Signing$\ $(PUF-based Cert-IBS with BLS) & 6.3 ms \\
		\midrule
        \rule{-2pt}{2ex}
		$\dagger$ values are averaged over 100 frames
    \end{tabular}
    \vspace{3mm}
	\caption{Running times for individual modules of application framework for the Zynq7010 SoC-based secure camera node}
    \label{table:runtimes}
\end{table}

\item \textbf{Storage.} Breakdown of the storage overhead incurred by the individual components of the proposed security mechanism is given by the second and third rows of Table~\ref{table:memory}. The size of the helper data corresponding to $160$ bit $sk$ and $128$ bit $k_E$ using the BCH codes is given by $1381$ bits and $1105$ bits, respectively (Table~\ref{tab:eval}). The node also needs to store $cert$ which requires $480$ bits ($60$ bytes). 

The Zynq7010 SoC stores three keys, which are used during the secure boot: a private $256$ bit key for AES256, a $256$ bit private key for HMAC-SHA$256$, and a $2048$ bit modular used by the RSA.  The size of a signed and encrypted partition is same as the unencrypted and unsigned one. Therefore, the storage overhead incurred by the secure boot functionality amounts to $320$ bytes. The Zynq7010 devices employ RSA for authentication of the boot partitions, which uses large key sizes ($2048$ bits). This can be replaced by a signature scheme based on elliptic curves such as elliptic curve digital signature algorithm (ECDSA), which provides the same level of security with a $256$ bit key and produces much smaller signatures. 

In comparison to memory consumed by the application logic, the storage overhead of the secure node mechanism is negligible. For example, the size of a double frame buffer (Table~\ref{table:memory}) used by the processing and communication tasks (Fig.~\ref{fig:videopath}) incur orders of magnitude greater storage overhead than the secure node mechanism.  

\begin{table}%[h]

	\centering
	\begin{tabular}{ll}
	\toprule
		\textbf{Module} & \textbf{Memory}\\
	\toprule
        \rule{-2pt}{3ex}
        Double Frame-Buffer & $2\times600$ kilobytes \\ 
		PUF Framework (Helper data) & $311$ bytes  \\
		%Frame Encyption (AES128) & 600kB \\ 
		Footage Signing ($cert$)            & $60$ bytes  \\
		Secure Boot  (RSA, AES, and HMAC\ keys)           & $320$ bytes  \\
		\midrule
	\end{tabular}
    \vspace{3mm}
    \caption{Memory consumption by the individual modules of the application framework for the Zynq7010 SoC-based secure camera node}
    \label{table:memory}
\end{table}	

\item \textbf{Hardware.} Hardware overhead incurs due to the RO PUF implementation in the FPGA part of the SoC. Table~\ref{tab:eval} gives the total hardware overhead for generating $160$ bit $sk$ and $128$ bit $k_E$ to $2492$ logic-gates, which is negligible as compared to the current state of the art security chip solutions.  

\item  \textbf{Communication.} Given an assisted living scenario, a camera with the proposed \textit{secure node} mechanism uploads encrypted-MACed-timestamped-signed footage, i.e., $\{C_i,\tau, \sigma\}_{i=i..N}$ to the edge or cloud storage server. The encrypted frame $C_i$ has the same size as the original image. A $256$ bit timestamp $\tau$ and a $160$ bit signature $\sigma$ are added to the footage for uploading. For a footage comprised of $N$ frames, the total communication overhead incurred on the secure camera node amounts only to $416$ bits, which amounts to $0.008 \%$ of a single frame size.
\end{itemize}

\subsection{Security and Privacy Properties}
\label{SecProp} 

This section discusses the security and privacy properties of the two schemes namely \textit{trusted sensing and privacy aware reporting} and \textit{secure node}. We also identify the limitations of our approaches and the additional measures that can be taken to over the limitations.

\subsubsection{Trusted Sensing and Privacy-Aware Reporting}
\label{Sec} 

The correctness of the trusted sensing and privacy-aware reporting scheme follows from the correctness of the PUF-based Cert-IBS scheme and the $P_\mathrm{NIWI}$ system. The system parameters $\Sigma$, $crs$ and $H$ are generated by the trusted authority. Since, any compromise to the trusted authority nullifies the trust and privacy guarantees, we emphasize that the \textit{offline} nature of the authority in our scheme greatly reduces the risk of compromise. The security and privacy properties of the privacy-aware trusted sensing scheme are as follows: 

\begin{itemize}

\item \textbf{Sensor-bound Secure Key Storage.} The on-chip PUF serves a secure key storage for the sensor. The key is generated from the hardware on sensor power up. During the off state, the key exist in form of unreadable variations introduced in the hardware by the CMOS manufacturing process. In comparison with a secure memory alternative, PUF offers a cost effective and more lightweight secure storage. Moreover, the PUF framework binds the key to the sensor hardware, which never leaves the sensor thereby minimizing the risk of a key compromise.

\item \textbf{Trusted Sensing.} Trusted sensing refers to the sensing techniques that ensure security guarantees about the integrity, authenticity and freshness of the sensed data. The scheme withstands the sensed data corruption attacks due to compromised OS running on mobile device. The OS receives sensors' readings accompanied by the commitments and proofs of $P_\mathrm{NIWI}$ from the root virtual machine (see Fig.~\ref{fig: SolScn1}). In order to inject fabricated data at the OS level, an attacker has to produce a valid $P_\mathrm{NIWI}$ proof of knowledge on a valid PUF-based Cert-IBS signature. A uf-cma secure PUF-based Cert-IBS implies that there is negligible probability that an attacker produces a valid PUF-based Cert-IBS signature. Further, the soundness of $P_\mathrm{NIWI}$ implies that it is impossible to generate a  valid proof without satisfying the PUF-based Cert-IBS verification equations. Further, any manipulation in the values of readings, commitments or proofs results in unsuccessful $P_\mathrm{NIWI}$ verification. Authenticity of each reading is ensured by signing the data with a unique, sensor-bound private key with-in the sensor. Freshness is ensured by including a time-stamp before signing the reading (sensed data attestation). In the participatory sensing scenario, freshness of sensed data is ensured by including a time-stamp, obtained from the smartphone GPS receiver, before the signature aggregation step.

\item \textbf{Privacy-Aware Reporting.} Anonymity of a user follows from the witness indistinguishability of $P_\mathrm{NIWI}$. Given $N$ mobile devices equipped with the privacy-aware trusted sensing scheme, contributing sensed data to an application server, each device is $N$-anonymous with respect to the server. The scheme provides CPA-anonymity~\cite{groth2007fully} since it does not provide the server oracle access to extract the witness from the proof using the extraction algorithm $X_{NI}$ of $P_\mathrm{NIWI}$. Since the server is assumed honest-but-curious threat to privacy, the CPA-anonymity suffices the privacy requirements. 

\item \textbf{Tamper Resistance.} Any physical tampering is detected due to the on-chip PUF since the PUF extracts a sensor fingerprint as a function of intrinsic details of the hardware. Any modification in hardware is detected as it results in generation of incorrect key.

\item \textbf{Identity Management.} Every sensor is assigned a unique physical identifier which could be used to  track and  monitor the state of a node, manage their access privileges in an all open network such as  IoT.

\end{itemize}

Next, we enlist some limitations of our scheme and identify some additional measures that may be appended with our scheme to overcome these limitations:
\begin{itemize}
    \item \textbf{Linkability and Location Privacy.} In order to ensure effective user privacy at the system level anonymization at the lower layers of the communication stack must be ensured since multiple reports can be trivially linked using the IP address or physical address of a user device (e.g., smartphone). Techniques such as mix networks, onion routing, IP rotation, and MAC address randomization can be leveraged for this purpose. For location privacy, appropriate location blurring technique such as cloaking, perturbation etc. must be used.
    
    \item \textbf{Event Faking.} Event faking attacks refer to capturing of sensed data from the undesirable environment. The proposed mechanism does not address these attacks. However, these can be thwarted by using statistical techniques on the server side.
    
\end{itemize}

\subsubsection{Secure Node}
Data and node security properties of the secure node scheme are as follows:

\begin{itemize}
\item \textbf{Secure Key Storage.} Secure storage of the keys is ensured by an on-chip PUF. On camera power up, the keys are generated from intrinsic variations of the hardware structure and are loaded into cache. When the camera device is off, the keys exist in form on unreadable variations introduced in the hardware by the CMOS manufacturing process. Compared to secure memory alternatives, PUF offers much cheaper secure storage.  

\item \textbf{Platform-bound Keys.} The PUF framework binds signing and encryption keys to the camera platform. The signing key never leaves the platform thereby minimizing the risk of the key compromise.  

\item \textbf{Data Nonrepudiation.} All video data leaving the camera carry integrity and authenticity guarantees. Authenticity is ensured by signing the video data using platform bound signing key. PUF-based Cert-IBS with BLS~\cite{boneh2001short} as underlying standard signature scheme is existentially unforgeable under chosen message attack (uf-cma secure), which is the standard security notion for a  digital signature algorithm. Any modification or fabrication of data during delivery or archival can be detected at the caretaker monitoring device. Spoofing using offline images can be addressed by using multiple sources for event detection, e.g., on-camera motion detection and sound detection. 

\item \textbf{Data Confidentiality.} Privacy of the monitored individual(s) is ensured by end-to-end encryption of each frame using AES128 algorithm.
\item[] \textbf{Access Authorization.} Secure key exchange and key storage on caretaker monitoring device ensure that only legitimate caretaker can access the decrypted video data. 

\item \textbf{Data Protection Lifetime.} Data is protected close to the source (sensor) and the security guarantees on the data remain valid for the entire lifetime of the data (i.e., during transmission, storage on cloud and delivery to monitoring device of caretaker). On the monitoring device, the data is consumed after successful verification of the security guarantees. 

\item \textbf{Camera Firmware Protection.} Each component of camera node's boot-chain (i.e., BootROM Code, FSBL, FPGA Bitstream, U-Boot, OS and Apps) is hashed, signed and encrypted using the HMAC-SHA256, RSA, and AES256 algorithms. At every boot-up, the integrity and authenticity of camera firmware is verified by through signature verification of each partition. Furthermore, the encryption of the boot partitions ensures unclonability of the camera firmware, since the decryption key exists within the security perimeter of the camera node. No hardware without the decryption key is capable of decrypting the firmware.  

\item \textbf{Physical Security.} First, on-chip PUF offers resistance against hardware tampering of the camera hardware. Since the PUF extracts device fingerprint as a function of intrinsic details of the hardware, hardware tampering is detected as it results in incorrect device fingerprint and keys. Second, in comparison with a TPM-based solution, where data is transferred from host processor to TPM chip (external to the host), where data protection mechanisms are applied. This results in exposed interface with unprotected data which can be tapped to bypass security mechanism. With incorporation of PUF in host SoC, these exposed interfaces are eliminated resulting in better physical security.
\end{itemize}

Next, we enlist some limitations of our scheme and identify some additional countermeasures that may be appended with our scheme to overcome these limitations:

\begin{itemize}

\item \textbf{Side Channel Attacks.}
First, although security keys are generated securely inside the SoC, key generation on every power-up opens up electromagnetic and power side channels. Analysis of the side channel information can result in partial recovery of keys at the hands of attackers. Approved effective techniques, masking (e.g, reversible process in which intermediate values of variables are randomized by masked with random numbers) and hiding (e.g., use of dual rail logic to flatten the data dependent leakage) can be leveraged to thwart side channel attacks. Second, data upload is triggered upon every event detection, the transmission pattern of the camera opens up another side-channel that leaks e.g., whether or not someone is at home. Transmitting dummy data at random intervals is a simple countermeasure that can be added to the system to mitigate this threat.

\item \textbf{Denial of Service.} Denial of service attacks by (i) the cloud, such as deleting the archived data, blocking the downloads or (ii) a third party, such as corrupting the video or control data in transit or storage are not addressed by this scheme. 

\item \textbf{Monitoring Device Security.}  The scheme uses symmetric key encryption (i.e., identical keys for encryption and decryption) since symmetric encryption is orders of magnitude faster and less power hungry than an asymmetric key encryption. However, symmetric key encryption requires secure key exchange between the camera and monitoring device and secure storage of key in the monitoring device.  Key exchange is done by the caretaker in a private space using a local interface, so the risk of key is relatively low. On monitoring device with untrusted software stack, virtualization, secure key vault or PUF can be used to securely store the key.   

\end{itemize}

To summarize, incorporation of trusted sensing and privacy-aware  reporting solution in the mobile devices can protect the participatory sensing applications against data pollution and personal privacy leakage attacks by ensuring three security guarantees: First, the sensed data remotely collected from ubiquitous commodity sensing devices carry integrity (i.e., no malicious manipulation of the data), authenticity (i.e., the data is originated from an authentic hardware sensor), and freshness (i.e., the data is recent and no adversary replayed an old message). Second, trusted sensing and privacy-aware  reporting scheme ensures anonymization and unlinkability of multiple submissions from the same device. Anonymization is given by the number of devices registered with or contributing to the application server. This prevents the server from creating personal profiles of the participating individuals. Incorporation of effective privacy protection mechanism would encourage increased user participants in the participatory sensing applications which is currently hindered due to privacy concerns. Third, the scheme improves the physical security of the sensors. On-chip PUF enables resistance against hardware tampering of the sensors. 

The trusted sensing and privacy-aware  reporting approach incurs $233$ bytes of storage, $8.77$ ms of latency and $1384$ logic-gates of hardware overhead on a sensor. To generalize, the latency overhead depends on the sensor clock and nature of the sensed data whereas the storage and hardware costs are fixed for all sensors. The communication overhead on the mobile device, for submitting $Q$ readings to an application server, is given by $(6+2Q)160+(4+2Q)320$ bits.

Participatory sensing application scenario was merely considered to illustrate the approach. However, the solution addresses all IoT applications that collect raw sensed data from commodity sensing devices to centralized server(s) and processing of the data takes place at the application server(s). 

Implementation of the secure node approach in visual monitoring applications thwarts the illegitimate data access and/or  manipulation and illegitimate control (over nodes and/or data) threats in these applications. The mechanism ensures integrity, authenticity, freshness, and confidentiality of the sensed data meanwhile allowing for processing of the data on the nodes. Threats to the personal privacy of the monitored individuals, either from a malicious application server or an unknown adversary, are addressed by ensuring data confidentiality and access authorization. The solution also protects the camera hardware and software  against hardware tampering, firmware manipulation, and firmware cloning attacks. This is particularly useful in cases where camera (or other sensor) nodes are deployed in unguarded open spaces. 

The secure node approach for assisted living scenario incurs $691$ bytes of storage, $27$ ms of latency, $2492$ logic-gates of hardware, and $416$ bits of communication overhead on a camera node, configured for $640\times480$ YUV422 format. The latency overhead depends on the node clock, type of sensor whereas the storage, hardware, and communication costs are same for any sensor node. 

The secure node ensures sensed data, personal privacy, and sensor node protection for the IoT applications. The security solution offers more flexibility as compared to trusted sensing and privacy-aware  reporting, as it additionally allows for the processing of sensed data on the sensor nodes. The solution is ideally suitable for visual monitoring applications however, it can be leveraged for other IoT applications that require processing of the sensed data locally on the sensor nodes.

\section{Conclusion} \label{conc}

Sensors are the largest and the most common source of data for the IoT-based smart services. Trustworthiness of sensed data and users' privacy are essential security requirements for these services. We presented lightweight and effective approach to protect the sensor nodes, sensed data and users' privacy for applications that require delivery of raw data to the server for processing and applications that require processing of sensed data locally on the sensor nodes. Both solutions ensure integrity, authenticity and freshness of sensed data, integrity of sensor hardware and usage privacy of the sensors. The scheme does not require any additional secure hardware and can be mapped on to the existing sensors' resources. Experimental evaluation of both solutions revealed that the proposed scheme incurs only insignificant latency, storage, hardware and communication overhead on the sensor nodes. Using the proposed sensors can effectively thwart sensed data pollution and privacy leakage attacks.

However, there are a number of open challenges that can be addressed as extension of this work. First open research problem is the privacy versus utility trade-off. During the privacy preservation process, the utility of the data diminishes as sensitive information such as the uniquely identifying information is removed, transformed, or distorted to achieve anonymity or confidentiality. Identifying the equilibrium between sensed data privacy and utility in IoT applications is an open challenge. Towards this goal, exploring other nuances of privacy between end-to-end encryption and full access could provide interesting insight. 

Second, the trusted sensor and secure node prototypes only demonstrate the proof of the concept however they were not optimized for performance. A number of performance improvements can be made in this regard. Considering the resource constraints of the sensors, further lightweight security primitives can be investigated. For instance, digital signature schemes require extensive resources that might not be economically feasible for many resource constrained IoT devices. Identification of lightweight security mechanisms to ensure integrity and authenticity guarantees is another open challenge in this regard.

\bibliographystyle{unsrtnat}
\bibliography{references.bib}

\begin{thebibliography}{56}
\providecommand{\natexlab}[1]{#1}
\providecommand{\url}[1]{\texttt{#1}}
\expandafter\ifx\csname urlstyle\endcsname\relax
  \providecommand{\doi}[1]{doi: #1}\else
  \providecommand{\doi}{doi: \begingroup \urlstyle{rm}\Url}\fi

\bibitem[Farrell and Barth(1999)]{farrell1999global}
J.~Farrell and M.~Barth.
\newblock \emph{{The Global Positioning System \& Inertial Navigation}},
  volume~61.
\newblock Mcgraw-Hill New York, NY, USA:, 1999.

\bibitem[Fueled()]{Fueled}
Fueled.
\newblock Paying with my phone, 2018.
\newblock \url{https://fueled.com/blog/10-ways-to-pay-with-your-smartphone/}.
\newblock Accessed 18 May 2018.

\bibitem[Clayton et~al.(2012)Clayton, Heaton, Chandy, Krause, Kohler, Bunn,
  Guy, Olson, Faulkner, and Cheng]{clayton2012community}
R.~Clayton, T.~Heaton, M.~Chandy, A.~Krause, M.~Kohler, J.~Bunn, R.~Guy,
  M.~Olson, M.~Faulkner, and M.~et~al. Cheng.
\newblock {Community seismic network}.
\newblock \emph{Annals of Geophysics}, 54\penalty0 (6):\penalty0 738--747,
  2012.

\bibitem[Carrapetta et~al.()Carrapetta, Youdale, Chow, and
  Sivaraman]{carrapetta2010haze}
J.~Carrapetta, N.~Youdale, A.~Chow, and V.~Sivaraman.
\newblock {Haze watch project, 2010}.
\newblock \url{http://www.pollution.ee.unsw.edu.au}.
\newblock Accessed 06 August 2018.

\bibitem[Das et~al.(2010)Das, Mohan, Padmanabhan, Ramjee, and
  Sharma]{das2010prism}
T.~Das, P.~Mohan, V.~Padmanabhan, R.~Ramjee, and A.~Sharma.
\newblock {PRISM: platform for remote sensing using smartphones}.
\newblock In \emph{Proceedings of the 8th International Conference on Mobile
  systems, applications, and services}, pages 63--76. ACM, 2010.

\bibitem[Schoenfeld et~al.(2004)Schoenfeld, Compton, Mead, Weiss, Sherfesee,
  Englund, and Mongeon]{schoenfeld2004remote}
M.~Schoenfeld, S.~Compton, H.~Mead, D.~Weiss, L.~Sherfesee, J.~Englund, and
  L.R. Mongeon.
\newblock Remote monitoring of implantable cardioverter defibrillators: A
  prospective analysis.
\newblock \emph{Pacing and clinical electrophysiology}, 27\penalty0
  (6p1):\penalty0 757--763, 2004.

\bibitem[eCAALYX(2015)]{ecaalyx}
eCAALYX.
\newblock {{Enhanced Complete Ambient Assisted Living Experiment}}, 2015.
\newblock URL \url{http://ecaalyx.org/ecaalyx.org/index.html}.
\newblock [Accessed: 17-May-2018].

\bibitem[Denning et~al.(2009)Denning, Andrew, Chaudhri, Hartung, Lester,
  Borriello, and Duncan]{denning2009balance}
T.~Denning, A.~Andrew, R.~Chaudhri, C.~Hartung, J.~Lester, G.~Borriello, and
  G.~Duncan.
\newblock {BALANCE: towards a usable pervasive wellness application with
  accurate activity inference}.
\newblock In \emph{Proceedings of the 10th Workshop on Mobile Computing Systems
  and Applications}, page~5. ACM, 2009.

\bibitem[Tesla()]{AD}
Tesla.
\newblock Self driving car demo, 2018.
\newblock \url{https://vimeo.com/192179727}.
\newblock Accessed 18 May 2018.

\bibitem[Hafner et~al.(2013)Hafner, Cunningham, Caminiti, and
  Del~Vecchio]{hafner2013cooperative}
M.R. Hafner, D.~Cunningham, L.~Caminiti, and D.~Del~Vecchio.
\newblock {Cooperative collision avoidance at intersections: Algorithms and
  experiments}.
\newblock \emph{IEEE Transactions on Intelligent Transportation Systems},
  14\penalty0 (3):\penalty0 1162--1175, 2013.

\bibitem[Lightner et~al.(2003)Lightner, Borrego, Myers, and
  Lowrey]{lightner2003wireless}
B.~Lightner, D.~Borrego, C.~Myers, and L.H. Lowrey.
\newblock Wireless diagnostic system and method for monitoring vehicles,
  October~21 2003.
\newblock US Patent 6,636,790.

\bibitem[Varaiya(1993)]{varaiya1993smart}
P.~Varaiya.
\newblock {Smart cars on smart roads: Problems of control}.
\newblock \emph{IEEE Transactions on automatic control}, 38\penalty0
  (2):\penalty0 195--207, 1993.

\bibitem[Cook et~al.(2003)Cook, Youngblood, Heierman, Gopalratnam, Rao, Litvin,
  and Khawaja]{cook2003mavhome}
D.~Cook, M.~Youngblood, E.~Heierman, K.~Gopalratnam, S.~Rao, A.~Litvin, and
  F.~Khawaja.
\newblock {MavHome: An agent-based smart home}.
\newblock In \emph{Proceedings of the First IEEE International Conference on
  Pervasive Computing and Communications}, pages 521--524. IEEE, 2003.

\bibitem[Ballesteros et~al.(2012)Ballesteros, Rahman, Carbunar, and
  Rishe]{ballesteros2012safe}
J.~Ballesteros, M.~Rahman, B.~Carbunar, and N.~Rishe.
\newblock {Safe cities: A participatory sensing approach}.
\newblock In \emph{Proceedings of the IEEE 37th Conference on Local Computer
  Networks}, pages 626--634. IEEE, 2012.

\bibitem[Atzori et~al.(2010)Atzori, Iera, and Morabito]{atzori2010internet}
L.~Atzori, A.~Iera, and G.~Morabito.
\newblock {The internet of things: A survey}.
\newblock \emph{Computer Networks}, 54\penalty0 (15):\penalty0 2787--2805,
  2010.

\bibitem[Saroiu and Wolman(2010)]{saroiu2010sensor}
S.~Saroiu and A.~Wolman.
\newblock {I am a sensor, and I approve this message}.
\newblock In \emph{Proceedings of the Eleventh Workshop on Mobile Computing
  Systems \& Applications}, pages 37--42. ACM, 2010.

\bibitem[G-DATA()]{Gdata2017}
G-DATA.
\newblock {{Mobile Malware Report 2017}}.
\newblock \url{https://www.gdatasoftware.com/mobile-internet-security-android}.
\newblock Accessed 17 May 2018.

\bibitem[eSecurity Planet()]{MasterKey}
eSecurity Planet.
\newblock {Inside the Bluebox Android Master Key Vulnerability, 2015}.
\newblock
  \url{http://www.esecurityplanet.com/mobile-security/inside-the-bluebox-android-master-key-vulnerability.html}.
\newblock Accessed 17 May 2018.

\bibitem[Security()]{FakeID}
Bluebox Security.
\newblock {Android FakeID Vulnerability, 2015}.
\newblock
  \url{https://www.blackhat.com/docs/us-14/materials/us-14-Forristal-Android-FakeID-Vulnerability-Walkthrough.pdf}.
\newblock Accessed 17 May 2018.

\bibitem[Liu et~al.(2012)Liu, Saroiu, Wolman, and Raj]{liu2012software}
H.~Liu, S.~Saroiu, A.~Wolman, and H.~Raj.
\newblock Software abstractions for trusted sensors.
\newblock In \emph{Proceedings of the 10th international conference on Mobile
  systems, applications, and services}, pages 365--378. ACM, 2012.

\bibitem[Winkler and Rinner(2014)]{Winkler_COMPSURV2014}
T.~Winkler and B.~Rinner.
\newblock {Security and Privacy Protection in Visual Sensor Networks: A
  Survey}.
\newblock \emph{ACM Computing Surveys}, 41\penalty0 (1):\penalty0 1--38, 2014.

\bibitem[Haider and Rinner(2017{\natexlab{a}})]{haidersecuring}
I.~Haider and B.~Rinner.
\newblock Securing cloud-based iot applications with trustworthy sensing.
\newblock In \emph{Proceedings of the 2nd EAI International Conference on
  Cloud, Networking for IoT}, pages 218--227. Springer, 2017{\natexlab{a}}.

\bibitem[Haider and Rinner(2017{\natexlab{b}})]{haider2017private}
I.~Haider and B.~Rinner.
\newblock {Private Space Monitoring with SoC-Based Smart Cameras}.
\newblock In \emph{Proceedings of the 14th International Conference on Mobile
  Ad-Hoc and Sensor Systems (MASS)}, pages 19--27. IEEE, 2017{\natexlab{b}}.

\bibitem[Dua et~al.(2009)Dua, Bulusu, Feng, and Hu]{dua2009towards}
A.~Dua, N.~Bulusu, W-C Feng, and W.~Hu.
\newblock Towards trustworthy participatory sensing.
\newblock In \emph{Proceedings of the 4th USENIX Conference on Hot topics in
  security}, pages 8--18. USENIX Association, 2009.

\bibitem[Dietrich and Winter(2009)]{dietrich2009implementation}
K.~Dietrich and J.~Winter.
\newblock Implementation aspects of mobile and embedded trusted computing.
\newblock In \emph{Proceedings of the International Conference on Trusted
  Computing}, pages 29--44. Springer, 2009.

\bibitem[Aaraj et~al.(2008)Aaraj, Raghunathan, and Jha]{aaraj2008analysis}
N.~Aaraj, A.~Raghunathan, and N.~Jha.
\newblock Analysis and design of a hardware/software trusted platform module
  for embedded systems.
\newblock \emph{ACM Transactions on Embedded Computing Systems (TECS)},
  8\penalty0 (1):\penalty0 1--31, 2008.

\bibitem[Winkler and Rinner(2010)]{winkler2010trustcam}
T.~Winkler and B.~Rinner.
\newblock {TrustCAM: Security and privacy-protection for an embedded smart
  camera based on trusted computing}.
\newblock In \emph{Proceedings of the IEEE Conference on Advanced Video and
  Signal-based Surveillance}, pages 593--600, 2010.

\bibitem[Winkler et~al.(2014)Winkler, Erd{\'e}lyi, and
  Rinner]{winkler2014trusteye}
T.~Winkler, A.~Erd{\'e}lyi, and B.~Rinner.
\newblock {TrustEYE. M4: Protecting the sensor--Not the camera}.
\newblock In \emph{Proceedings of the IEEE Conference on Advanced Video and
  Signal-based Surveillance}, pages 159--164, 2014.

\bibitem[Erd{\'e}lyi et~al.(2014)Erd{\'e}lyi, Barat, Valet, Winkler, and
  Rinner]{erdelyi2014avss}
A.~Erd{\'e}lyi, T.~Barat, P.~Valet, T.~Winkler, and B.~Rinner.
\newblock {Adaptive Cartooning for Privacy Protection in Camera Networks}.
\newblock In \emph{Proceedings of the IEEE Conference on Advanced Video and
  Signal-based Surveillance}, pages 44--49, 2014.

\bibitem[Potkonjak et~al.(2010)Potkonjak, Meguerdichian, and
  Wong]{potkonjak2010trusted}
M.~Potkonjak, S.~Meguerdichian, and J.~Wong.
\newblock {Trusted sensors and remote sensing}.
\newblock In \emph{Proceedings of the 9th Annual IEEE Conference on Sensors},
  pages 1104--1007. IEEE, 2010.

\bibitem[Cornelius et~al.(2008)Cornelius, Kapadia, Kotz, Peebles, Shin, and
  Triandopoulos]{cornelius2008anonysense}
C.~Cornelius, A.~Kapadia, D.~Kotz, D.~Peebles, M.~Shin, and N.~Triandopoulos.
\newblock {Anonysense: Privacy-aware people-centric sensing}.
\newblock In \emph{Proceedings of the 6th International Conference on Mobile
  systems, Applications, and Services}, pages 211--224. ACM, 2008.

\bibitem[De~Cristofaro and Soriente(2011)]{de2011short}
E.~De~Cristofaro and C.~Soriente.
\newblock Short paper: Pepsi---privacy-enhanced participatory sensing
  infrastructure.
\newblock In \emph{Proceedings of the fourth ACM Conference on Wireless Network
  Security}, pages 23--28. ACM, 2011.

\bibitem[Dimitriou et~al.(2012)Dimitriou, Krontiris, and
  Sabouri]{dimitriou2012pepper}
T.~Dimitriou, I.~Krontiris, and A.~Sabouri.
\newblock Pepper: a querier's privacy enhancing protocol for participatory
  sensing.
\newblock In \emph{Proceedings of the International Conference on Security and
  Privacy in Mobile Information and Communication Systems}, pages 93--106.
  Springer, 2012.

\bibitem[Rosenfeld et~al.(2010)Rosenfeld, Gavas, and
  Karri]{rosenfeld2010sensor}
K.~Rosenfeld, E.~Gavas, and R.~Karri.
\newblock {Sensor physical unclonable functions}.
\newblock In \emph{Proceedings of the International Symposium on
  Hardware-Oriented Security and Trust (HOST)}. IEEE, 2010.

\bibitem[Cao et~al.(2015)Cao, Zhang, Zalivaka, Chang, and Chen]{cao2015cmos}
Y.~Cao, L.~Zhang, S.~Zalivaka, C-H Chang, and S.~Chen.
\newblock {CMOS image sensor based physical unclonable function for coherent
  sensor-level authentication}.
\newblock \emph{IEEE Transactions on Circuits and Systems}, 62\penalty0
  (11):\penalty0 2629--2640, 2015.

\bibitem[Wired()]{iPhonetracking}
Wired.
\newblock iphone tracks your every move, and there's a map for that, 2011.
\newblock \url{https://www.wired.com/2011/04/iphone-tracks/}.
\newblock Accessed 14 June 2018.

\bibitem[Groth and Sahai(2012)]{groth2012efficient}
J.~Groth and A.~Sahai.
\newblock Efficient noninteractive proof systems for bilinear groups.
\newblock \emph{SIAM Journal on Computing}, 41\penalty0 (5):\penalty0
  1193--1232, 2012.

\bibitem[Chen et~al.(2008)Chen, Garfinkel, Lewis, Subrahmanyam, Waldspurger,
  Boneh, Dwoskin, and Ports]{chen2008overshadow}
X.~Chen, T.~Garfinkel, C.~Lewis, P.~Subrahmanyam, C.~Waldspurger, D.~Boneh,
  J.~Dwoskin, and D.~Ports.
\newblock {Overshadow: A virtualization-based approach to retrofitting
  protection in commodity operating systems}.
\newblock In \emph{Proceedings of the 13th International Conference on
  Architectural Support for Programming Languages and Operating Systems
  (ASPLOS)}, pages 2--13. ACM, 2008.

\bibitem[Brakensiek et~al.(2008)Brakensiek, Dr{\"o}ge, Botteck, H{\"a}rtig, and
  Lackorzynski]{brakensiek2008virtualization}
J.~Brakensiek, A.~Dr{\"o}ge, M.~Botteck, H.~H{\"a}rtig, and A.~Lackorzynski.
\newblock Virtualization as an enabler for security in mobile devices.
\newblock In \emph{Proceedings of the 1st Workshop on Isolation and Integration
  in Embedded Systems}, pages 17--22. ACM, 2008.

\bibitem[Tuyls and Batina(2006)]{tuyls2006rfid}
P.~Tuyls and L.~Batina.
\newblock {RFID-tags for Anti-Counterfeiting}.
\newblock In \emph{Proceedings of Topics in Cryptology - Cryptographers Track
  at the RSA Conference}, pages 115--131. Springer, 2006.

\bibitem[Bellare et~al.(2009)Bellare, Namprempre, and
  Neven]{bellare2009security}
M.~Bellare, C.~Namprempre, and G.~Neven.
\newblock Security proofs for identity-based identification and signature
  schemes.
\newblock \emph{Journal of Cryptology}, 22\penalty0 (1):\penalty0 1--61, 2009.

\bibitem[Boneh et~al.(2004)Boneh, Boyen, and Shacham]{boneh2004short}
D.~Boneh, X.~Boyen, and H.~Shacham.
\newblock Short group signatures.
\newblock In \emph{Proceedings of the Annual International Cryptology
  Conference}, pages 41--55. Springer, 2004.

\bibitem[Treece()]{CPUvFPGA}
B.~Treece.
\newblock {CPU or FPGA for image processing: Which is best?, 2017}.
\newblock
  \url{https://www.vision-systems.com/articles/print/volume-22/issue-8/features/cpu-or-fpga-for-image-processing-which-is-best.html}.
\newblock Accessed 18 May 2018.

\bibitem[Collins et~al.(2000)Collins, Lipton, Kanade, Fujiyoshi, Duggins, Tsin,
  Tolliver, Enomoto, Hasegawa, Burt, and Wixson]{collins2000system}
R.~Collins, A.~Lipton, T.~Kanade, H.~Fujiyoshi, D.~Duggins, Y.~Tsin,
  D.~Tolliver, N.~Enomoto, O.~Hasegawa, P.~Burt, and L.~Wixson.
\newblock A system for video surveillance and monitoring.
\newblock Technical Report CMU-RI-TR-00-12, Robotics Institute, Carnegie Mellon
  University, 2000.

\bibitem[Boneh et~al.(2001)Boneh, Lynn, and Shacham]{boneh2001short}
D.~Boneh, B.~Lynn, and H.~Shacham.
\newblock {Short signatures from the Weil pairing}.
\newblock In \emph{Proceedings of the International Conference on the Theory
  and Application of Cryptology and Information Security}, pages 514--532.
  Springer, 2001.

\bibitem[Sanders(2013)]{sanders2013secure}
L.~Sanders.
\newblock {Secure boot of Zynq-7000 all-programmable SoC}.
\newblock \emph{Application Note XAPP1175, Xilinx}, 2013.

\bibitem[Maes(2012)]{roel2012physically}
R.~Maes.
\newblock \emph{{Physically unclonable functions: Constructions, properties and
  applications}}.
\newblock PhD thesis, University of KU Leuven, 2012.

\bibitem[Rajendran et~al.(2016)Rajendran, Tang, and
  Karri]{rajendran2016securing}
J.~Rajendran, J.~Tang, and R.~Karri.
\newblock {Securing pressure measurements using SensorPUFs}.
\newblock In \emph{Proceedings of the International Symposium on Circuits and
  Systems (ISCAS)}, pages 1330--1333. IEEE, 2016.

\bibitem[Kodytek and Lorencz(2015)]{kodytek2015newPUF}
F.~Kodytek and R.~Lorencz.
\newblock {A Design of Ring Oscillator Based PUF on FPGA}.
\newblock In \emph{Proceedings of the 18th International Symposium on Design
  and Diagnostics of Electronic Circuits \& Systems}, pages 37--42. IEEE, 2015.

\bibitem[Guajardo et~al.(2007)Guajardo, Kumar, Schrijen, and
  Tuyls]{guajardo2007physical}
J.~Guajardo, S.~Kumar, G.-J. Schrijen, and P.~Tuyls.
\newblock {Physical unclonable functions and public-key crypto for FPGA IP
  protection}.
\newblock In \emph{Proceedings of the International Conference on Field
  Programmable Logic and Applications}, pages 189--195. IEEE, 2007.

\bibitem[Lynn(2016)]{pbc}
B.~Lynn.
\newblock {The Pairing-Based Cryptography Library}, 2016.
\newblock URL \url{https://crypto.stanford.edu/pbc/}.
\newblock [Accessed: 17-May-2018].

\bibitem[Barbulescu et~al.(2013)Barbulescu, Gaudry, Joux, and
  Thom{\'e}]{cryptoeprint:2013:400}
R.~Barbulescu, P.~Gaudry, A.~Joux, and E.~Thom{\'e}.
\newblock A quasi-polynomial algorithm for discrete logarithm in finite fields
  of small characteristic.
\newblock Cryptology ePrint Archive, Report 2013/400, 2013.
\newblock URL \url{http://eprint.iacr.org/2013/400}.

\bibitem[Barreto and Naehrig(2005)]{barreto2005pairing}
P.~Barreto and M.~Naehrig.
\newblock Pairing-friendly elliptic curves of prime order.
\newblock In \emph{Proceedings of the International Workshop on Selected Areas
  in Cryptography}, pages 319--331. Springer, 2005.

\bibitem[Google()]{GoogleSV}
Google.
\newblock {Street View, 2016}.
\newblock \url{https://www.google.com/streetview/camera-loans/#eligibility}.
\newblock Accessed 17 May 2018.

\bibitem[Lab(2015)]{WikiRome}
MIT Senseable~City Lab.
\newblock {WikiCity Rome}, 2015.
\newblock URL \url{http://senseable.mit.edu/wikicity/rome/}.
\newblock [Accessed: 17-May-2018].

\bibitem[Groth(2007)]{groth2007fully}
J.~Groth.
\newblock Fully anonymous group signatures without random oracles.
\newblock In \emph{Proceedings of the International Conference on the Theory
  and Application of Cryptology and Information Security}, pages 164--180.
  Springer, 2007.

\end{thebibliography}
\end{document}